\documentclass[prd,superscriptaddress,floatfix,preprintnumbers,altaffilletter,twocolumn,nofootinbib]{revtex4}
\usepackage{amsmath}
\usepackage{amssymb}
\usepackage{cancel}
\usepackage{color}
\usepackage{epsfig}
\usepackage[mathscr]{euscript}
\usepackage{graphicx}
\usepackage{longtable}
\usepackage{placeins}
\usepackage{rotating}
\usepackage{subfigure}
\usepackage{units}
\usepackage{url}
\usepackage{verbatim}
\usepackage{wasysym}

\begin{document}

\newcommand*{\rom}[1]{\uppercase\expandafter{\romannumeral #1\relax}}

\newcommand*\dif{\mathop{}\!\mathrm{d}}

\pagenumbering{arabic}

\title{Second Einstein Telescope Mock Data and Science Challenge: Low Frequency Binary Neutron Star Data Analysis}

\author{Duncan Meacher}
\email{Duncan.Meacher@ligo.org}
\affiliation{Laboratoire Artemis, Universit\'{e} C\^{o}te d'Azur, CNRS, Observatoire C\^{o}te d'Azur, Bd de l'Observatoire,  BP 4229, 06304, Nice Cedex 4, France}
\affiliation{Dept. of Physics, Pennsylvania State University, 104 Davey Lab, University Park, PA 16802, USA}

\author{Kipp Cannon}
\affiliation{ Canadian Institute for Theoretical Astrophysics, 60 St. George Street, University of Toronto, Toronto, Ontario, M5S 3H8, Canada}

\author{Chad Hanna}
\affiliation{Dept. of Physics, Pennsylvania State University, 104 Davey Lab, University Park, PA 16802, USA}

\author{Tania Regimbau}
\affiliation{Laboratoire Artemis, Universit\'{e} C\^{o}te d'Azur, CNRS, Observatoire C\^{o}te d'Azur, Bd de l'Observatoire,  BP 4229, 06304, Nice Cedex 4, France}

\author{B. S. Sathyaprakash}
\affiliation{School of Physics and Astronomy, Cardiff University, 5 The Parade, Cardiff, CF24 3AA, UK}

\date{\today}

\begin{abstract}
The Einstein Telescope is a conceived third generation gravitational-wave detector that is envisioned to be an order of magnitude more sensitive than advanced LIGO, Virgo and Kagra, which would be able to detect gravitational-wave signals from the coalescence of compact objects with waveforms starting as low as 1Hz.
With this level of sensitivity, we expect to detect sources at cosmological distances.
In this paper we introduce an improved method for the generation of mock data and analyse it with a new low latency compact binary search pipeline called \texttt{gstlal}.
We present the results from this analysis with a focus on low frequency analysis of binary neutron stars.
Despite compact binary coalescence signals lasting hours in the Einstein Telescope sensitivity band when starting at 5 Hz, we show that we are able to discern various overlapping signals from one another.
We also determine the detection efficiency for each of the analysis runs conducted and show a proof of concept method for estimating the number signals as a function of redshift.
Finally, we show that our ability to recover the signal parameters has improved by an order of magnitude when compared to the results of the first mock data and science challenge.
For binary neutron stars we are able to recover the total mass and chirp mass to within 0.5\% and 0.05\%, respectively.
\end{abstract}

\maketitle


\section{Introduction}
\label{sec:intro}

Second generation gravitational-wave (GW) detectors, aLIGO~\cite{cqg.32.074001.15} and AdVirgo~\cite{cqg.32.024001.15}, are planned to improve the sensitivity over first generation detectors, LIGO~\cite{rpp.72.076901.09} and Virgo~\cite{aip.794.307.05} by an order of magnitude. 
aLIGO has recently begun operations and AdVirgo is currently in the commissioning stage with plans to join operations in 2016. 
It is expected that the first direct detection of gravitational waves will be made before the end of this decade.

The Einstein Telescope (ET) is a conceived third generation gravitational-wave detector that is currently in the design stage~\cite{cqg.27.194002.10} and is planned to be operational after $\sim$ 2025.
This detector will have an improvement in sensitivity by an order of magnitude over that of the second generation detectors that will allow for the detection of a large number of GW signals from a variety of processes, out to large distances.
These include, but are not limited to, events such as the formation of neutron stars or black holes from core collapse supernovae~\cite{prd.72.084001.05,prd.73.104024.06,mnras.398.293.09,mnrasl.409.L132.10}, rotating neutron stars~\cite{aap.376.381.01,prd.86.104007.12}, and the merger of compact binary systems~\cite{prd.84.084004.11,prd.84.124037.11}.

ET is expected to yield a significant number of detections and the interpretation of the results will allow us to answer questions about astrophysics, cosmology and fundamental interactions~\cite{cqg.29.124013.12}.
In order to prepare and test our ability to extract valuable information from the data, we initiated a series of mock data and science challenges (MDSCs), with increasing degrees of sophistication and complexity with each subsequent challenge.
These challenges consist of first simulating ET data that includes a population of sources expected to be detectable via different astrophysical models.
This is then analysed with a variety of current data analysis algorithms, each searching for a specific signal type contained within the data.
Unlike advanced detectors, ET data is expected to be dominated by many overlapping signals which increases the complexity of the data analysis. 
An important goal of the MDSC is to test the ability of different analysis algorithms in efficiently detecting signals and discriminating different signal populations.
Finally we consider the interpretation of these results to investigate different areas of astrophysics and cosmology.

For the first ET MDSC~\cite{prd.86.122001.12}, we produced one month of mock data containing simulated Gaussian coloured noise, produced using a plausible ET noise power spectral density (PSD), and the GW signals from a set of compact binary coalescence (CBC), in this case a population of binary neutron stars (BNS) in the redshift range $z\in$[0, 6].
Using a modified version of the LIGO/Virgo data analysis pipeline \texttt{ihope}~\cite{prd.79.122001.09,prd.80.047101.09,prd.82.102001.10,prd.85.082002.12}, which was the main matched filtering analysis pipeline during the initial detector era, we showed that it is possible to employ the use of a matched filtering algorithm to search for GW signals when there is a large amount of overlap of their waveforms.
Using this pipeline we were also able to recover the observed chirp mass ($\mathcal{M}_z$) and observed total mass ($M_z$) of the injected signals with an error of less than 1\% and 5\% respectively\footnote{
The observed mass parameters, $M_z$ and $\mathcal{M}_z$, differ from the intrinsic parameters, $M$ and $\mathcal{M}$, by a factor of (1+$z$), due to the redshifting of the GW frequencies from the expansion of the Universe, which is the equivalent of observing heavier masses.
These are denoted with a subscript $z$, such that $\mathcal{M}_z \equiv \mathcal{M}(1+z)$.}.
We also analysed the data with the standard isotropic cross-correlation statistic and measured the amplitude of an astrophysical stochastic GW background (SGWB)~\cite{prd.59.102001.99,prd.79.062002.09,raap.11.369.11} created by the population of background BNS signals with an accuracy better than 5\%.
Finally, we were able to verify the existence of a \textit{null stream}, created by the closed loop detector layout which results in the complete cancelling of GW signals and gives an acceptable estimate of the noise PSD of the detectors.
By subtracting the null stream from the data, we showed that we could recover the expected shape of the PSD of the astrophysical SGWB.

After the success of the first challenge, we extended our data generation package to conduct a second MDSC.
The second ET MDSC contains a larger selection of sources over that of the first, including BNS, neutron star-black holes (NSBH), binary black holes (BBH), binary intermediate mass black holes (IMBH)~\cite{grg.43.485.11} as well as several burst sources.
In the second MDSC we have taken the intrinsic mass distributions and time delays, the time between the formation and merger of the binary systems, from the population synthesis code \texttt{StarTrack}~\cite{apj.572.407.02,apjs.174.223.08,apjl.715.L138.10,apj.759.52.12}, as opposed to selecting the component masses from a Gaussian distribution in the first MDSC.
With this mock data set several investigations have been carried out, each focusing on a different scientific aspect of the MDSC. 
The first of these investigations, on the measurement of a SGWB from astrophysical sources, has already been completed~\cite{prd.89.084046.14}, while others are ongoing.

In this paper we investigate the application of a new low-latency matched filtering analysis pipeline, \texttt{gstlal}~\cite{prd.82.044025.10,prd.83.084053.11,apj.784.136.12,prd.88.024025.13}, which is built using \texttt{gstreamer} multimedia processing technology.
The analysis will be run multiple times, searching for low mass systems, using a low frequency cut-off of 25Hz, 10Hz and 5Hz, on both the main mock data set as well as a noise only data set that is used to make estimates of the background.
The 25Hz and 10Hz runs will be conducted on the full data set while the 5Hz analysis will be run on 10\% of the data.
This is due to the fact that starting at 5Hz, there are more templates produced for the analysis and the waveform for low mass systems will be of the order of a few hours long, both of which significantly increases the computational cost of the analysis.

Once the analyses have been run, we compare the list of detections that are reported in each of the three ET detectors against the list of injected signals.
Using a small window in both coalescence time $(t_c)$ and the observed (redshifted) chirp mass ($\mathcal{M}_z$) we produce a list of matched detections.
We will then make a comparison of the recovered detection parameters ($t_c$, $\mathcal{M}_z$ and $M_z$) against the true injected parameters.

The rest of this paper is divided into the following sections.
In Section~\ref{sec:MockData} we introduce the methods by which we produce the mock data used for this investigation.
In Section~\ref{sec:analysis} we discuss the analysis methods that are used as well as our reasons for choosing a new analysis pipeline.
In Section~\ref{sec:results} we present our results from the analysis runs that are conducted, with a focus on both event detection and parameter measurements.
In Section~\ref{sec:futuredev} we highlight possible areas that can be investigated in future MDSCs.
Finally in section~\ref{sec:conclusion} we discuss the results shown in the last section and make a conclusion to this investigation.


\section{Mock Data}
\label{sec:MockData}

In this section we describe how we go about generating the ET mock data used in this investigation.
Here we use the same data generation package as was used in the first ET MDSC~\cite{prd.86.122001.12}, which has since been updated to simulate more sources~\cite{prd.89.084046.14,prd.92.063002.15}.
We first explain the generation of the coloured noise and then we introduce and describe each of the steps that are used to simulate the GW inspiral signals that are injected into the noise.
For this we describe how the cosmological model and star formation rate (SFR) are used to determine the rate of coalescence of compact binary objects as a function of redshift and how the signal parameters are selected as well as the waveform models used in the simulation.

\subsection{Simulation of the Noise}

\begin{figure*}
\includegraphics[width=0.49\textwidth]{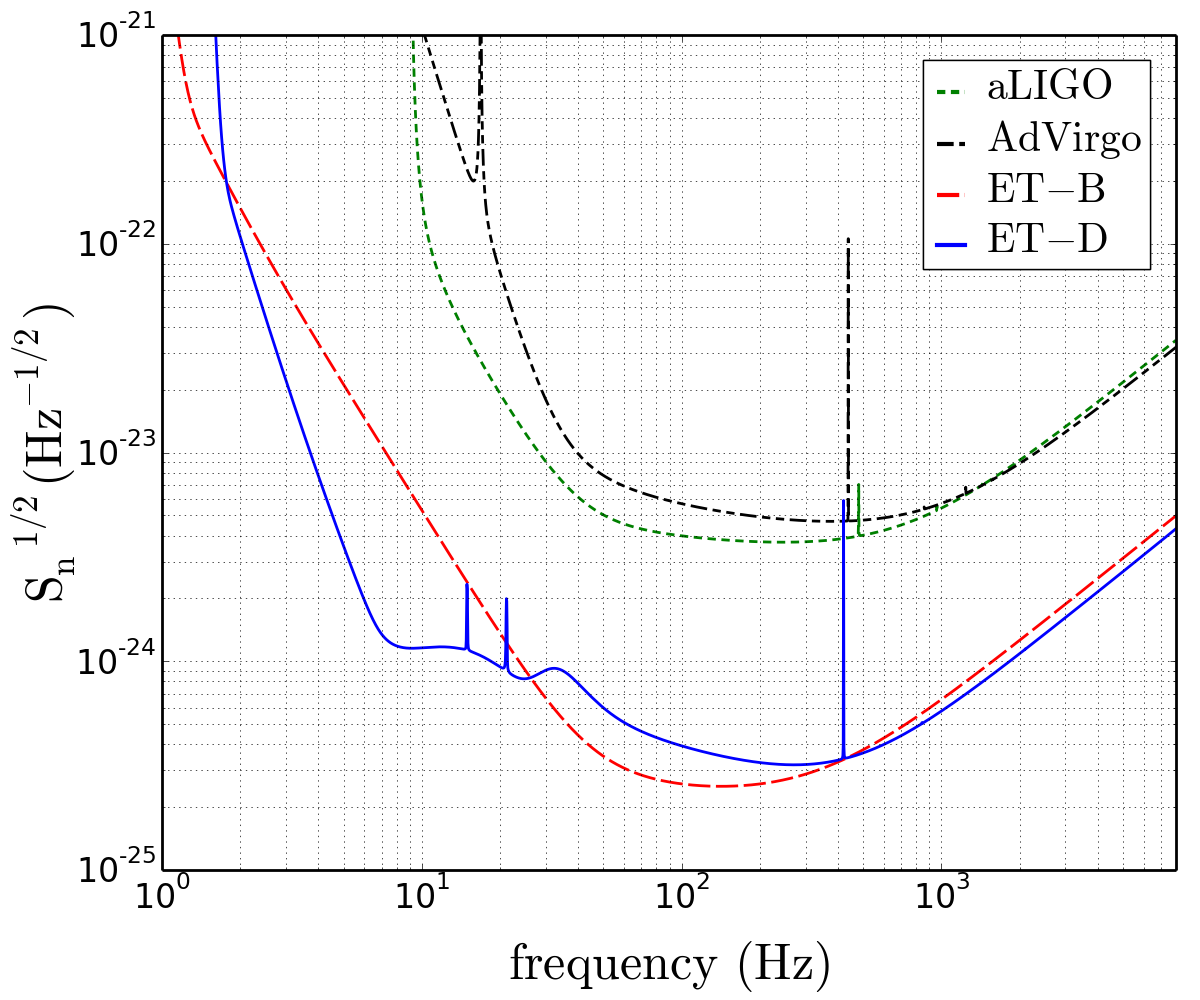}
\includegraphics[width=0.49\textwidth]{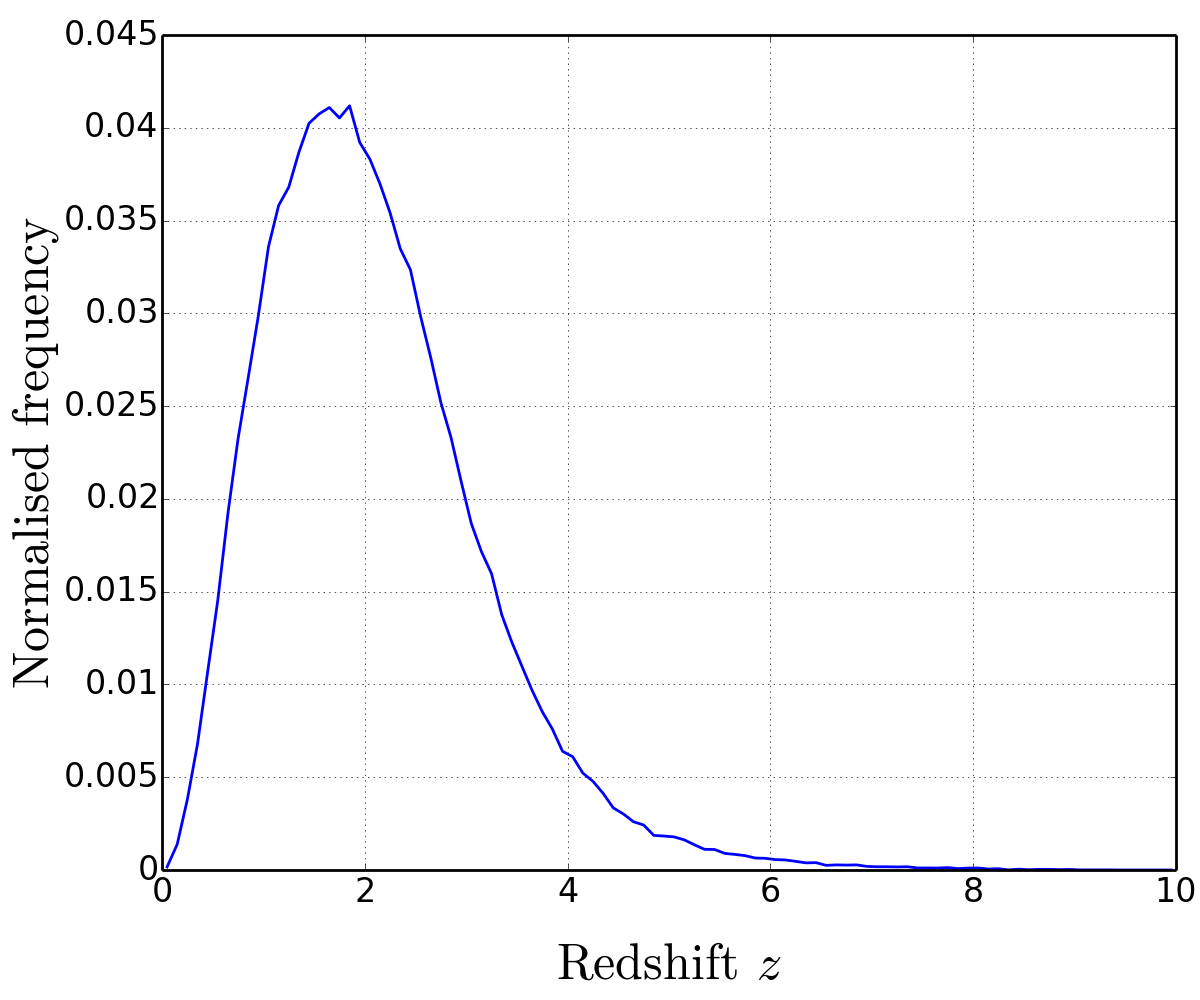}
\caption{\emph{Left}--- Projected design noise power spectral density for advanced LIGO (dot-dot green), advanced Virgo (dot-dashed black), ET-B (dashed red) and ET-D (solid blue).
\emph{Right}--- Normalised distribution of the redshift for all BNS events, using redshift bins of size $\Delta z = 0.1$, as provided by \texttt{StarTrack}.}
\label{fig:noise}
\end{figure*}

The current design of the Einstein Telescope is envisioned to consist of three independent V-shaped Michelson interferometers with 60 degree opening angles, arranged in a triangle configuration, and placed underground to reduce the influence of seismic noise~\cite{cqg.26.085012.09,ijmpd.22.1330010.13}.
Here we make the assumption that there will be no instrumental or environmental correlated noise between the detectors so that the noise is simulated independently for each of the three ET detectors, E1, E2 and E3~\cite{prd.87.123009.13,prd.90.023013.14}.
This is done by generating a Gaussian time series that has a mean of zero and unit variance.
This time series is then Fourier transformed into the frequency domain, coloured with the noise PSD of the ET detector, and then inverse Fourier transformed back into the time domain.
In order to remove any potential discontinuities between adjacent data segments, we gradually taper away the noise spectral density to zero at frequencies above 4096Hz and below 5Hz, which we set as the low frequency cut-off for the generation of the noise and GW signals.
For this MDSC, we consider the sensitivity given by ET-D rather than ET-B that was used in the first MDSC, as shown in the left-hand plot in Fig.~\ref{fig:noise}.
ET-B is a simpler design with just one interferometer in each V of the equilateral triangle but due to high stored power it suffers from enhanced radiation pressure noise at lower frequencies.
ET-D is a design that includes two interferometers in each V (a high-frequency, high-power interferometer to mitigate photon shot noise and a low-frequency, low-power, cryogenics interferometer to mitigate thermal noise) and achieves a very good high-frequency sensitivity without compromising on low-frequency sensitivity.

\subsection{Simulation of the GW signals from BNS}

We employ the use of Monte Carlo (MC) simulation techniques for the generation of the mock data. 
The process that we use to generate the various parameters is very similar to that used in the first ET MDSC~\cite{prd.86.122001.12}, except here we take the intrinsic mass distribution of the component masses, $m_1$ and $m_2$, and the time delay, $t_d$, i.e. the interval between the formation of a binary and its eventual merger, from the stellar evolution code \texttt{StarTrack}~\cite{apj.572.407.02,apjs.174.223.08,apjl.715.L138.10,apj.759.52.12}.
As was done in the first MDSC, we adopt a $\Lambda$CDM cosmological model with the Hubble parameter $H_0 = 70$ km s$^{-1}$ Mpc$^{-1}$, $\Omega_m = 0.3$, and $\Omega_\Lambda = 0.7$ and the SFR of~\cite{apj.651.142.06}.
We first consider the coalescence rate for BNS per unit volume, as a function of redshift

\begin{equation}
\dot{\rho}_c(z, t_d) \propto  \frac{\dot{\rho}_\ast (z_f(z,t_d))}{1+z_f(z,t_d)} , \;\; \text{with} \;\; \dot{\rho}_c(0) = \dot{\rho} _0 ,
\end{equation}

\noindent where $z$ is the redshift of the source at the point of coalescence, $z_f$ is the redshift of the source at the point at which the binary formed, $\dot{\rho}_\ast$ is the SFR and $\dot{\rho}_0$ is the local coalescence rate. 
A factor of $(1+z_f)^{-1}$ is used to convert the rate from the source's frame of reference to the observer's frame of reference. 

The redshifts $z$ and $z_f$ are connected to each other via the delay time, $t_d$, which is the total time that it takes between the initial formation of the binary system, through its evolution into a compact binary and finally the merging time to the point of coalescence due to the emission of gravitational radiation using

\begin{equation}
t_d = \frac{1}{H_0} \int_z^{z_f} \frac{\dif{z}'}{(1+{z}')E({z}')},
\end{equation}

\noindent where

\begin{equation} \label{eq:EcosParam}
E(z) = \sqrt{\Omega_m(1+z)^3 + \Omega_\Lambda}.
\end{equation}

The coalescence rate per redshift bin is given by

\begin{equation} \label{eq:CoalescenceRate}
\frac{\dif R}{\dif z}(z,t_d) = \dot{\rho}_c(z,t_d) \frac{\dif V}{\dif z}(z),
\end{equation}

\noindent where $\dif V / \dif z$ is the comoving volume element given by

\begin{equation}
\frac{\dif V}{\dif z}(z) = 4 \pi \frac{c}{H_0} \frac{r^2(z)}{E(z)},
\end{equation}

\noindent where $c$ is the speed of light in vacuum and $r(z)$, the proper distance, is given by

\begin{equation}
r(z) = \frac{c}{H_0} \int_0^z \frac{\dif{z}'}{E({z}')} .
\end{equation}

The average time between the arrival of events, which we define as $\lambda$,  is given by taking the inverse of the coalescence rate, Eq.~(\ref{eq:CoalescenceRate}), integrating over all redshifts

\begin{equation}
\lambda = \left[ \int_0^{z_\mathrm{max}} \frac{\dif R}{\dif z} (z,t_d) \dif z \right]^{-1}.
\end{equation}

\begin{figure*}
\includegraphics[width=0.49\textwidth]{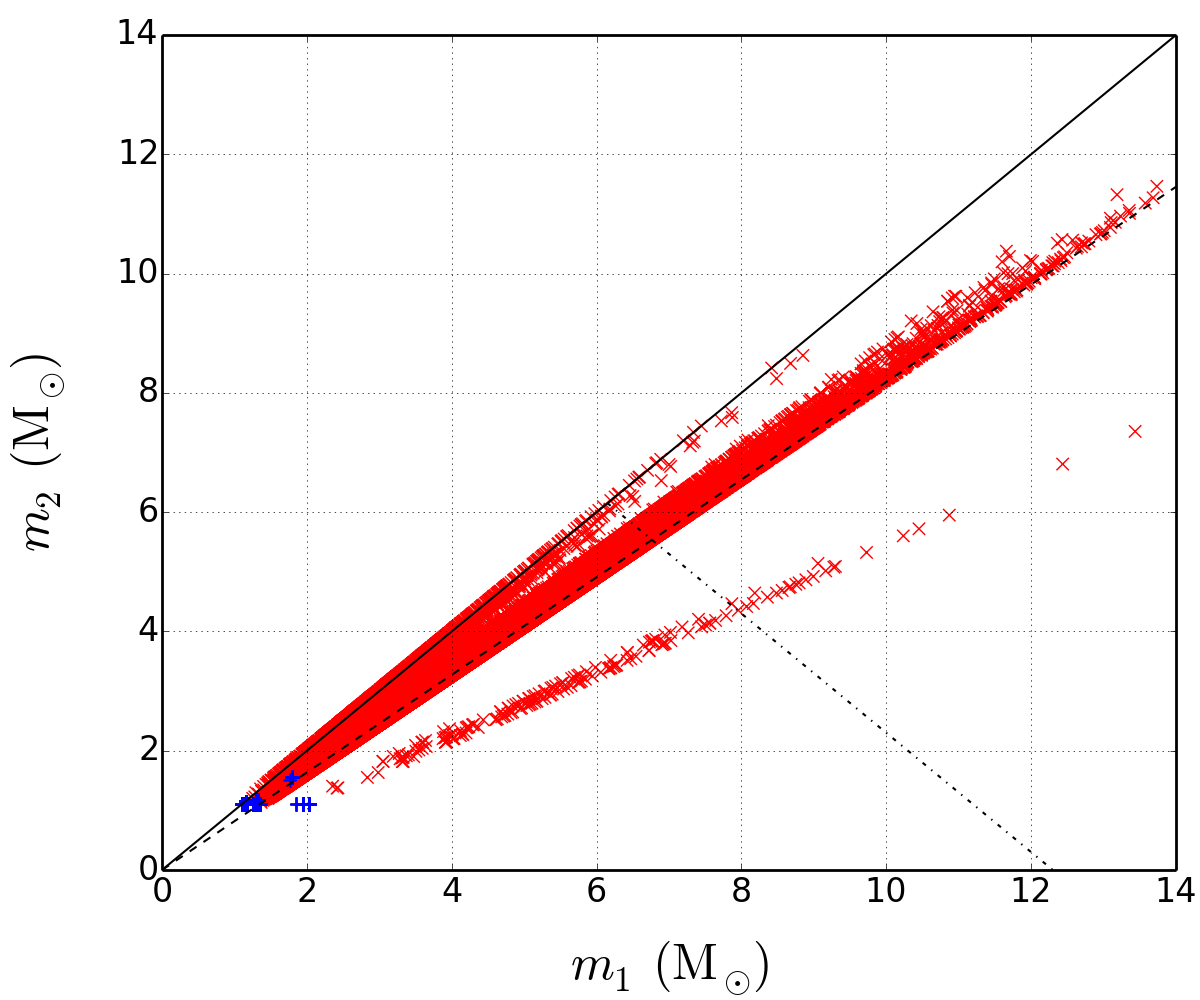}
\includegraphics[width=0.49\textwidth]{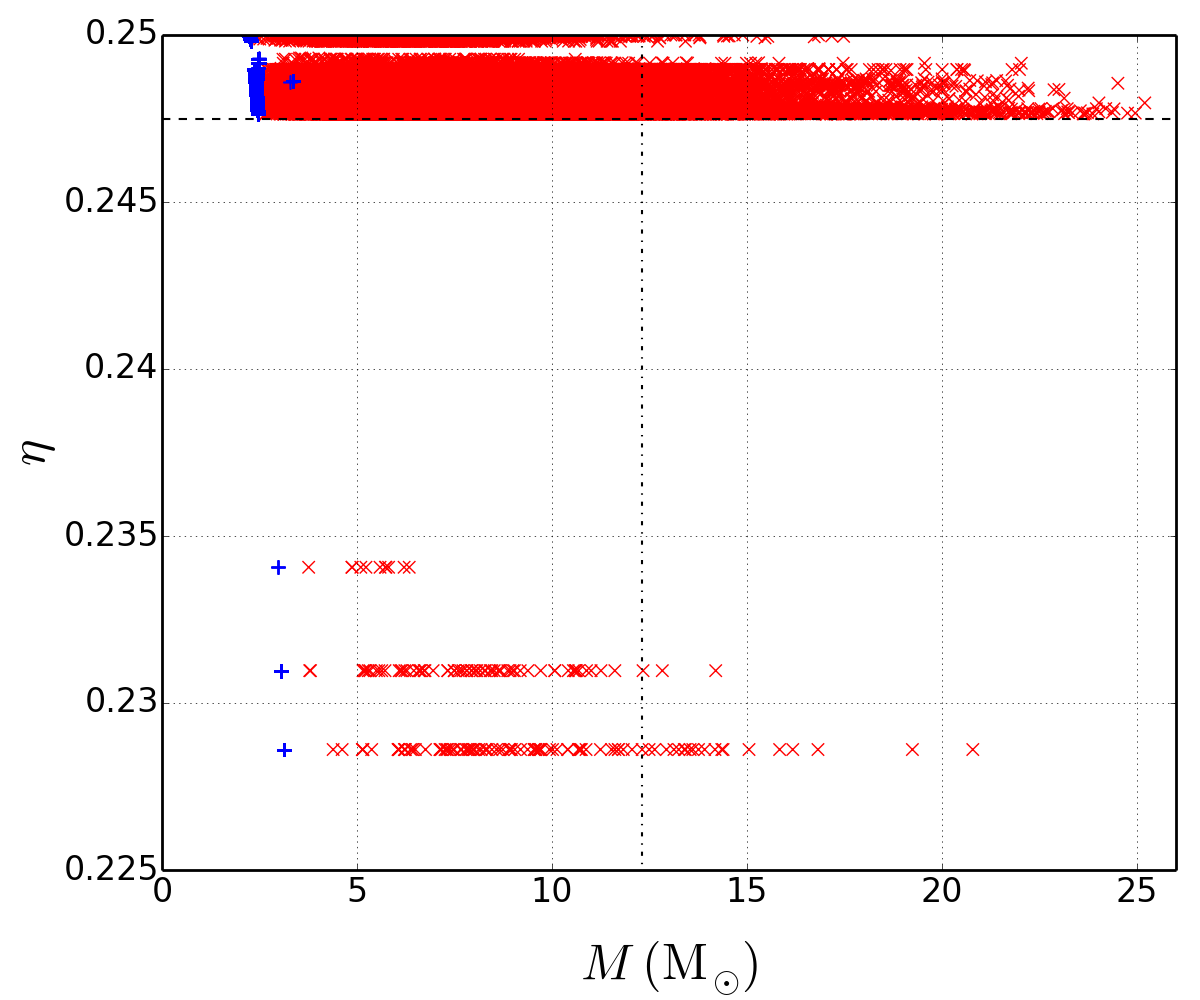}
\caption{\emph{Left}--- Injected masses, $m_1$ and $m_2$, where $m_1 \geq m_2$.
The blue + are the intrinsic masses and the red $\times$ are the observed (redshifted) masses, for 128244 BNS, as given by \texttt{StarTrack}.
The diagonal solid black line represents equal masses with $\eta$~=~0.25, where $\eta = m_1m_2 / M^2$, the diagonal dashed line represents $\eta$~=~0.2475, and the dot-dashed line represents a total mass of 12.3$\mathrm{M}_\odot$.
\emph{Right}--- Injected total mass, $M$, against symmetric mass ratio, $\eta$, where the blue + are the intrinsic values and the red $\times$ are the observed (redshifted) values, for 128244 BNS, as given by \texttt{StarTrack}.
The dashed horizontal line represents $\eta$~=~0.2475 and the dot-dashed vertical line represents a total mass of 12.3$\mathrm{M}_\odot$.}
\label{fig:InjMass}
\end{figure*}

Once we have a value for the average waiting time between events we then produce the parameters for each CBC source as follows:

\begin{itemize}
\item The arrival time, $t_c$, of injection $i$ is selected assuming a Poisson distribution, where the difference in arrival time, $\tau = t_c^i - t_c^{i-1}$, is drawn from an exponential distribution $P(\tau) = \exp(-\tau / \lambda)$.

\item The average time between all events is set to $\lambda$~=~20~s, which is comparable to the realistic rate given in~\cite{cqg.27.173001.10} where different coalescence rates for BNS, NSBH, BBH and IMBH are taken into account\footnote{The original data sets as presented in~\cite{prd.89.084046.14} consisted of a year's worth of data that had an average time between all injection of $\lambda$~=~200~s, provided from Table 3 in~\cite{aap.574.A58.15} using the BZ model.
In order to reduce the computational cost of running the analysis with a very low cut-off frequency we have reduced the amount of data by a factor of 10 while increasing the coalescence rate by the same factor.
This means that the same injections are present within both sets while the time of arrival between successive events has decreased resulting in more overlap of the waveforms.
It has already been shown in~\cite{prd.86.122001.12} that this overlap does not affect the ability of a matched filtering algorithm to detect overlapping signals.} .
This gives a total of 159,302 events which are split up into the following proportions: 80.47\% BNS (128,244), 2\% NSBH (3190), 12.46\%  BBH (19,766), provided from Table 3 in~\cite{aap.574.A58.15}, and 5.07\% IMBH (8102).

\item The binary's component masses, $m_1$ and $m_2$, shown in Fig.~\ref{fig:InjMass}, and the time delay, $t_d$, are selected from a list of compact binaries generated by \texttt{StarTrack}.
For the given delay time and a particular model for the cosmic SFR, we construct a redshift probability distribution, $p(z, t_d)$, by normalising the coalescence rate in the interval $z$ = [0, 10], where

\begin{equation}
p(z, t_d) = \lambda \dfrac{\dif R}{\dif z} (z,t_d) .
\end{equation}

In the right-hand plot of Fig.~\ref{fig:noise} we show the normalised redshift distribution for BNS, produced by using redshift bins of size $\Delta z = 0.1$. 

\item The sky position, $\hat{\Omega}$, the cosine of the inclination angle, $\iota$, the polarization angle, $\psi$, and the phase at the coalescence, $\phi_0$, are selected from uniform distributions.

\item The two GW polarisation amplitudes, $h_+(t)$ and $h_\times(t)$, and the antenna response functions to the two polarisations for each of the three ET detectors, $F^A_+(t,\hat{\Omega},\psi)$ and $F^A_\times(t,\hat{\Omega},\psi)$, where $A$ = 1, 2, 3 is the index representing one of the three ET detectors, are then calculated.
The detector responses

\begin{equation}
h^A(t) = F^A_+(t,\hat{\Omega},\psi) h_+(t) + F^A_\times(t,\hat{\Omega},\psi) h_\times(t) , 
\end{equation}

are then added to the detector output time series for E1, E2 and E3, where the modulation of the signal due to the rotation of Earth is taken into account.
In this MDSC we have chosen to use the TaylorT4 waveforms \cite{prd.80.084043.09}, which is accurate to 3.5 post-Newtonian order~\cite{lrr.17.2.14}, in phase and the most dominant lowest post-Newtonian order term in amplitude, for the generation of the BNS and NSBH signals. 
For the BBH signals we choose the EOBNRv2 waveforms \cite{prd.87.082004.13} that includes the merger and quasi-normal ring down phases of the signal, and it is accurate to $4^{th}$ post Newtonian order in phase and lowest order in amplitude~\cite{prd.80.084043.09}.
\end{itemize}

For the sake of testing and to determine the number of background detections we might expect to have, we have also produced a second, noise only data set that is produced with the same Gaussian noise as the main data set.

\section{Analysis}
\label{sec:analysis}

The analysis method used here to search for the CBC signals is generally the same as was used in the first MDSC though we are now using a newly developed pipeline, \texttt{gstlal}.
This is a coincident analysis pipeline where the data streams from each of the separate detector's are analysed individually via matched filtering with the use of a large bank of templates.
The template bank is produced using a TaylorF2 waveform~\cite{prd.62.084036.00}, which is generated in the frequency domain to the second post Newtonian order and terminates at the frequency of the last stable circular orbit, where $f_\mathrm{lsco}~\simeq~\dfrac{c^3}{6^{3/2} \pi G M_z}$.
This waveform generator is selected as it is relatively fast to generate (compared to the TaylorT4 waveform) and reduces the computational cost of the analysis which is performed in the frequency domain.
The analysis produces a list of matched \emph{triggers} that exceed a given SNR threshold, $\rho_\mathrm{T}$; each trigger is a list that contains the SNR and the parameters of the template that produced the trigger, such as the epoch of merger and component masses of the  binary.
These are then checked against triggers from the other two detectors for coincidence.
Any double or triple coincident triggers that result from the same template are then reported as potential GW detections though in this investigation we only consider the results from triple coincident events.

\subsection{Analysis stages}

The different stages for this analysis pipeline are described here:

\begin{itemize}
\item Estimation of PSD:
The \texttt{gstlal} analysis estimates the noise PSD as function of time during filtering. 
The method is a modified version of Welch's method~\cite{itae.15.70.67} with two main differences. 
First, each periodogram is derived from choosing the geometric mean of the last 7 periodograms and second, the periodograms are weighted averages that weigh the present periodogram slightly more than the past ones.  
The result is a PSD estimate with an effective average over a few hundred seconds with 1/16 Hz resolution.

\item Generation of template bank:
A bank of GW inspiral signals are produced that are used to search the data. 
This bank needs to cover the full mass parameter range that is being considered.
Because we know the mass distributions of the signals being injected we are able to tailor the mass parameter limits that are used to generate the template banks in order to cover the full range of masses whilst keeping the number of templates produced to a minimum.
A new template bank is generated for each search that is conducted, with the mass parameter ranges given in Table~\ref{tab:searches}.

\item Matched filtering:
This is implemented with the LLOID (Low Latency Online Inspiral Detection) method, which uses singular value decomposition (SVD) to compress the waveform parameter space and multi-rate time domain filtering~\cite{apj.784.136.12}.  
It provides the same result as standard matched filtering~\cite{prd.85.122006.12} to within $< 1\%$.
The matched filtering of each SVD bank against each detector data stream produces an SNR time series $\rho(t)$.

\item Trigger generation: 
As templates are filtered against data streams, if any SNR time series passes a threshold value, $\rho_\mathrm{T}$, then it is considered as a trigger.
Generally, using a lower SNR threshold value is better as it allows for the possibility of detecting weaker signals but it also results in an increase in the number of triggers produced from background noise.
Here we set the single detector threshold to be SNR = 4 as this is the lowest we can go without having a trigger rate that becomes difficult to deal with.

\item Coincidence between detectors: 
Triggers from different detectors are then compared against each other.
Any that are coincident in time, within a 5 ms window to account for small time delays for the time of flight between detectors, and have the same masses, are considered as either double or triple coincident triggers.
The SNR for a network of detectors is given by 

\begin{equation}
\rho^2 = \sum_A \rho^2_A\,.
\end{equation}

For triple coincident triggers this gives a minimum SNR of $\sim$~6.928.

\item Clustering of triggers:
The list of double and triple coincident triggers is then clustered, where any coincident events that occur within a 4 second time window of a coincident events with a higher SNR are deleted.
This is done as the same event will be detected by multiple templates, some with a certain degree of mismatch in the signal parameters.
This results in the reporting of the best matched template.

\end{itemize}

The output of \texttt{gstlal}, containing all clustered triple coincident triggers, are then compared against the list of injections in order to ``match" any potential detections.
For this we apply a time and chirp mass window to each detection and if an injection is found within this two dimensional window then we determine it to be a found injection.
If two injections are found within the same two dimensional window then the injections with the smallest redshift is assumed to be the more likely event.
The chirp mass is selected because, as was found in the first MDSC and as is shown later, it is better constrained than the total mass by the analysis.
Here a time window of $\pm$100~ms and a chirp mass window of 1\% of the observed chirp mass for BNS is used.

\begin{table*}
\caption{\label{tab:searches} A list of all searches carried out in this investigation.
The first column gives the identity of the search.
The second column indicates if the analysis was run on the noise only or main data set.
The third column gives the low frequency cut-off used for the analysis run.
The fourth column gives the total length of the search in seconds.
The fifth column gives the total mass range used for the search.
The sixth column gives the symmetric mass ratio range used.
The final column gives the total number of templates produced given the previous search parameters before the singular value decomposition is applied.}
\begin{center}
\begin{tabular}{ c c c c c c c}
\hline \hline
Search & Data & f$_\mathrm{min}$ (Hz) & length (s) & $M_\mathrm{total}$ range (M$_\odot$) & $\eta$ range & N$_\mathrm{templates}$ \\ \hline
1   & Noise + Signals & 25 & 3072000 & 2.6 - 12.3 & 0.2475 - 0.25 & 3603 \\ 
2   & Noise + Signals &10  & 3072000 & 2.6 - 12.3 & 0.2475 - 0.25 & 25252 \\
3   & Noise + Signals & 5   & 307200   & 2.6 - 12.3 & 0.2475 - 0.25 & 87054 \\ \hline
4   & Noise                & 25 & 3072000 & 2.6 - 12.3 & 0.2475 - 0.25 & 3647 \\
5   & Noise                & 10 & 3072000 & 2.6 - 12.3 & 0.2475 - 0.25 & 26173 \\
6   & Noise                & 5   & 307200   & 2.6 - 12.3 & 0.2475 - 0.25 & 89495 \\
\hline \hline
\end{tabular}
\end{center}
\end{table*}

\subsection{Searches}

Compared to the standard advanced detector searches there are several differences that we implement here.
The first is low frequency cut-off used to produce the signal templates.
Advanced detector will only be sensitive down to $\sim$ 20Hz for the first couple of years of operations, eventually reduced to $\sim$ 10Hz when the detectors begin to operate at the design sensitivity~\cite{ObsScenario}.
Starting at these frequencies, low mass systems will have waveform lengths of only a few minutes to tens of minutes.
When considering ET, which is sensitive down to frequencies as low as 1-3Hz, depending on the final design configuration, signal templates can be of the order of hours to several days in length.
In this investigation will focus on the application of different low frequency cut-offs where we run three searches using the same template mass range but using different $f_\mathrm{min}$.
We use a low frequency cut-off of 25Hz and 10Hz where we analyse the full mock data, and then analyse 10\% of the data at 5Hz.
We select one analysis run at 25Hz so that we can make a direct comparison to the results from the first MDSC and we choose to only analyse 10\% of the data at 5Hz because of the high computational cost associated with this analysis.
At this starting frequency with the injected masses shown in Fig.~\ref{fig:InjMass}, the template waveform lengths are already several hours long.
Because of this we also impose a cut-off at a redshift of $z$ = 0.2, below which our search templates will not be sensitive.
Instead we make the assumption that we have a detection efficiency of 100\%.
After this point, the signals are redshifted by a factor of $(1+z)$ by a significant fraction so that the signal wavelengths become computationally manageable.
For these searches we set a minimum component mass of 1.3M$_\odot$, minimum total mass of 2.6M$_\odot$, a maximum component mass of 6.75M$_\odot$ and a maximum total mass of 12.3M$_\odot$ with a minimum symmetric mass ratio of $\eta = m_1m_2 / M^2 = 0.2475$.
This minimum symmetric mass ratio is chosen to be as high as possible to reduce the number of templates being generated whilst still including most of the population of BNS, as can be seen in the right-hand plot of Fig.~\ref{fig:InjMass}.
Already at this $\eta_\mathrm{min}$ we produce $\sim87000$ templates when starting at 5Hz.
All the search parameters are displayed in Table~\ref{tab:searches}.

All three analysis runs are repeated on the noise only data sets in order to obtain an estimate on the number of background triggers one would expect in the main data set.
From these results an SNR threshold value is set with which to make a cut on all triggers in the main data sets.
For this we select the SNR equal to the 100th loudest events for the 25Hz and 10Hz runs, and the 10th loudest event for the 5Hz run. 
At present there is no method for determining an estimate for the false alarm probability with ET and so the 100th (10th) loudest noise event is selected as it will cover most of the population of background noise events whilst avoiding statistical fluctuations which produce louder SNR events that may skew the background estimate.
The results of this are presented in Table~\ref{tab:detections}.


\section{Results}
\label{sec:results}

In this section we present the results from all the analysis runs carried out as part of this investigation, which is divided into four sub-sections.
The first shows the number of detections made for each analysis run and the second details the detection efficiency.
In the third we explore a proof of concept method for estimating the number of injected signals as a function of redshift and the fourth presents the accuracy with which we are able to recover the injection parameters.

\begin{table*}
\caption{\label{tab:detections} A list of the number of triggers and detections produced for different SNR threshold values used with each search.
The first column gives the identity of the search.
The second column gives the number of triggers produced when analysing the noise only data set.
The third column give the SNR of the 100th loudest event obtained from the noise only data set.
The fourth and fifth columns gives the total number of triggers and matched detections produced when no SNR threshold cut is is applied.
The sixth and seventh columns gives the total number of triggers and matched detections with an SNR larger than that of the 100th loudest event from the noise only data set.
For the two right-hand columns the number in the brackets indicates the remaining percentage of triggers and detections compared to when the lowest SNR threshold cut is used.}
\begin{center}
\begin{tabular}{ c | c c | c c | c c }
\hline \hline
 & \multicolumn{2}{c|}{Noise} & \multicolumn{2}{c|}{$\rho_\mathrm{T} = 6.9$} & \multicolumn{2}{c}{$\rho_\mathrm{T} =$ SNR (100th loudest noise event)}  \\ \hline
Search & N$_\mathrm{triggers}$ & SNR (100th loudest) & N$_\mathrm{triggers}$ & N$_\mathrm{detections}$ & N$_\mathrm{triggers}$ & N$_\mathrm{detections}$ \\ \hline
1 & 74323   & 8.655 & 82322   & 5708 & 5670 (6.89\%) & 4713 (82.57\%) \\ 
2 & 291319 & 8.904 & 341747 & 9956 & 15590 (4.56\%) & 8138 (81.74\%)  \\ 
3 & 45183    & 8.964\footnote{Due to the reduced amount of data that has been analysed at 5Hz, we have selected the SNR of the 10th loudest event from the noise only analysis run.}  & 63709   & 1242 & 7320 (11.49\%) & 1095 (88.19\%)  \\ 
\hline \hline
\end{tabular}
\end{center}
\end{table*}

\subsection{\texttt{gstlal} analysis: Impact of the lower frequency cut-off on detection efficiency}

The results for the different analysis runs with different low frequency cut-offs are summarized in Table~\ref{tab:detections}.
Here the first column gives the search identity, the second column gives the number of triggers that were produced when analysing the noise only data set, and the third column gives the SNR of the 100th (10th) loudest event.
The fourth and fifth columns give the total number of triggers and resulting number of matched detections that are made with the smallest possible network SNR threshold of 6.9.
The sixth  and seventh columns again show the number of triggers and matched detections corresponding to an SNR threshold, $\rho_\mathrm{T}$, equal to the 100th (10th) loudest event from the noise only data set.
The number in the brackets for the two right-hand  columns indicates the fractional number of triggers or matched detections that remain when a higher SNR threshold is used as compared to the case of smallest SNR theshold.

\begin{figure}
\includegraphics[width=0.45\textwidth]{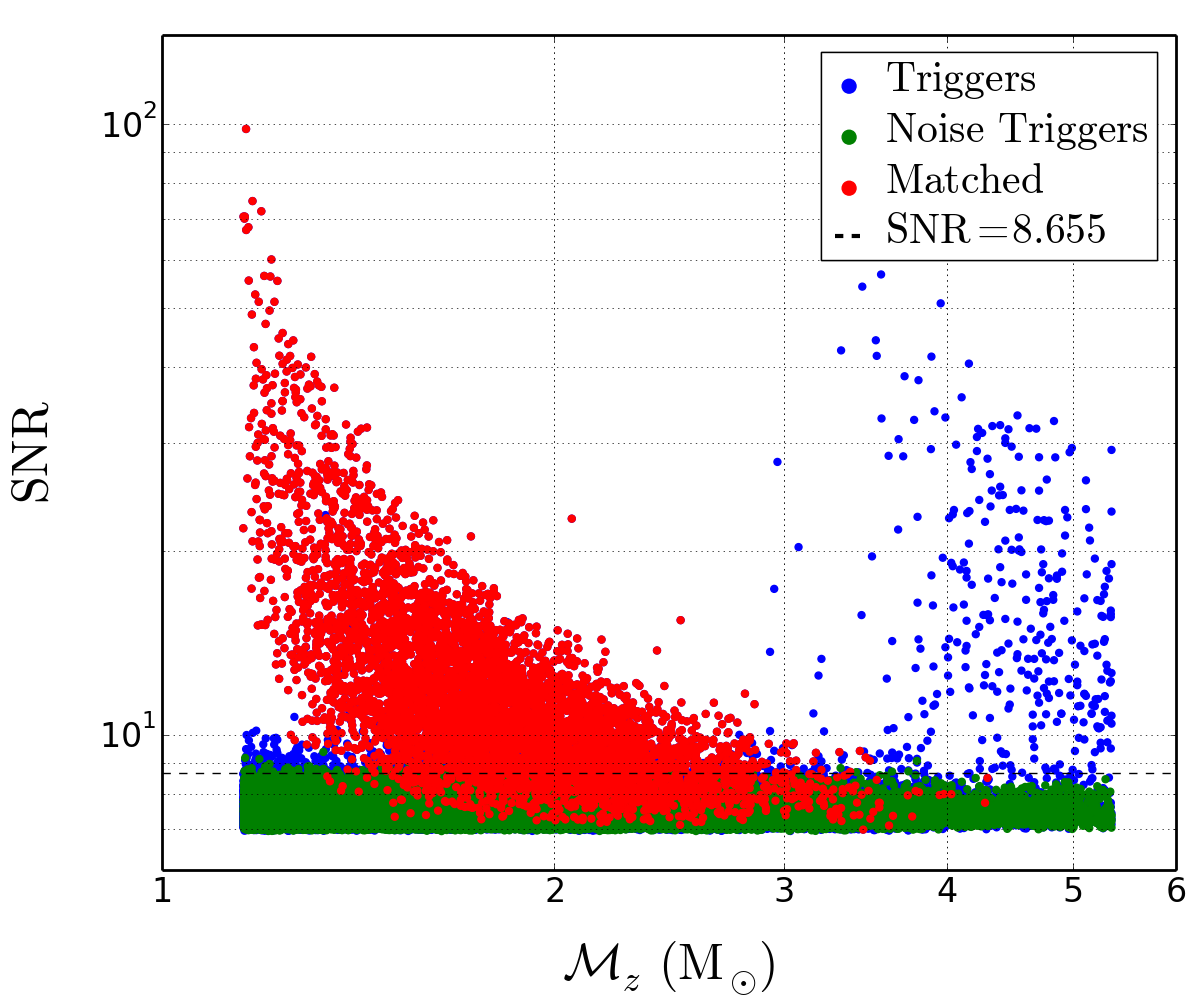} \\
\includegraphics[width=0.45\textwidth]{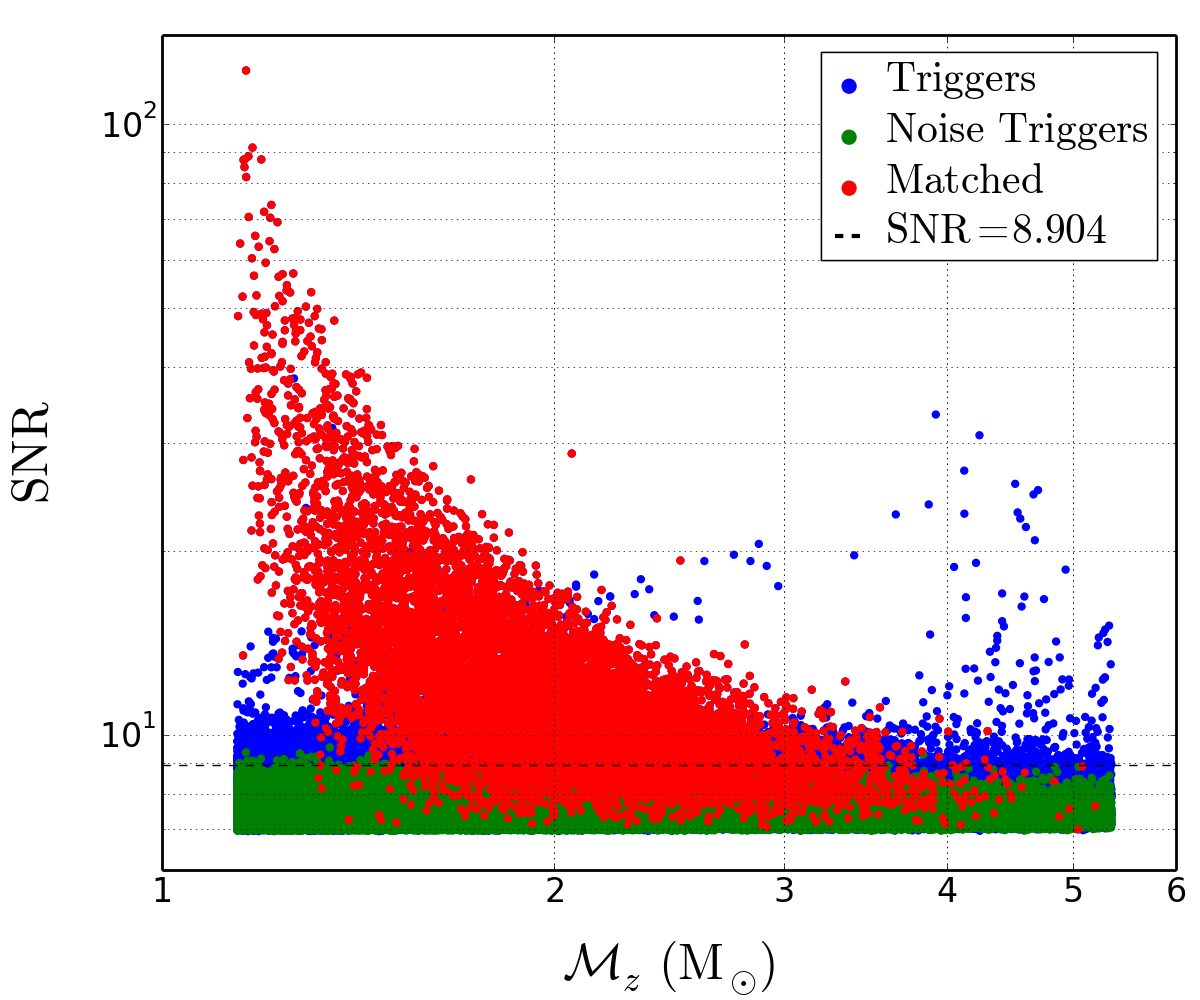} \\
\includegraphics[width=0.45\textwidth]{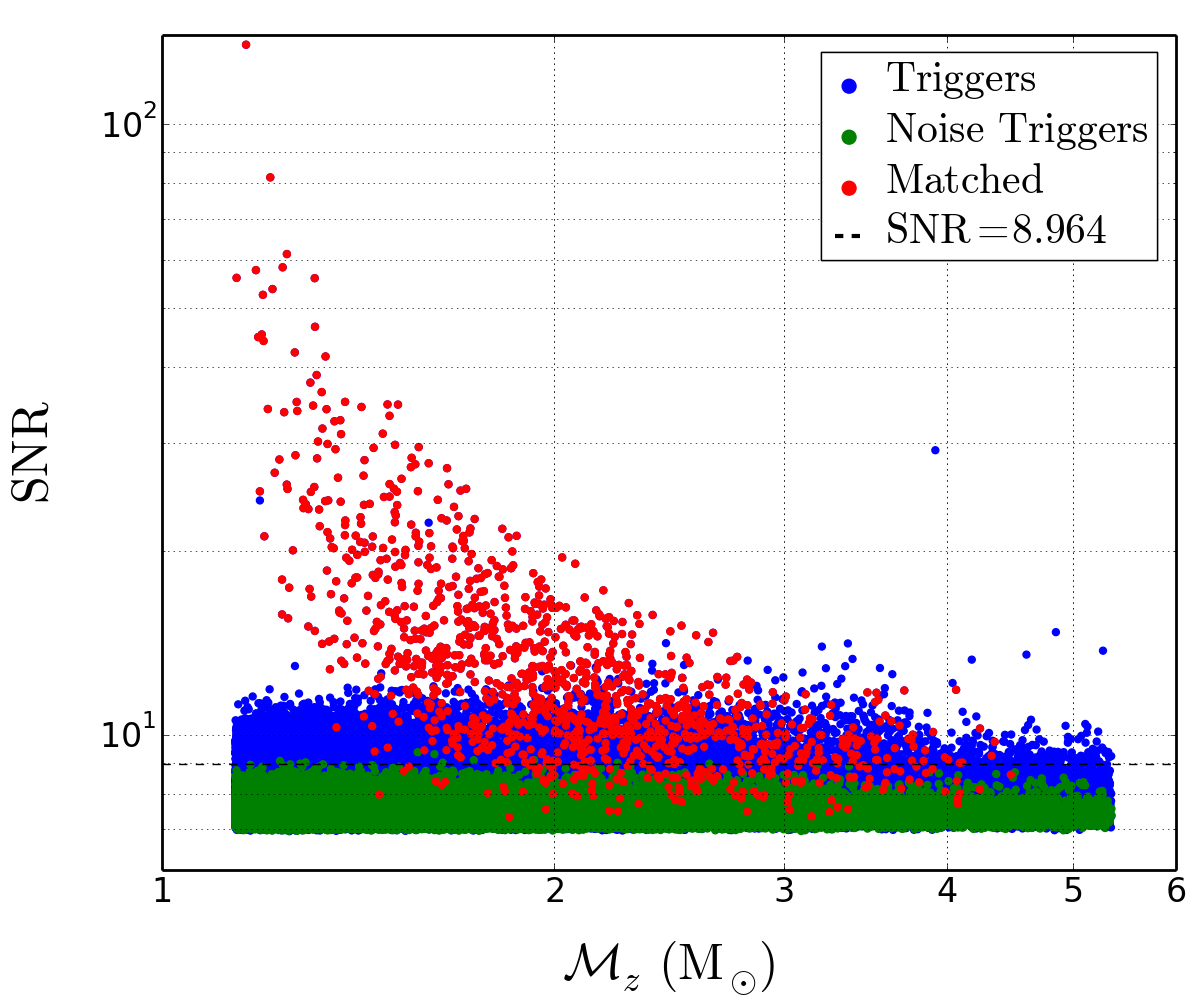}
\caption{Scatter plots of SNR against the observed chirp mass for the three different low frequency cut-offs used in the analysis with 25Hz (top), 10Hz (middle) and 5Hz (bottom).
All triggers produced from the analysis of the main data set are shown in blue, with the triggers produced from the analysis of the noise only data set shown in green.
Any of the triggers from the main data set that are then matched to an injection are then plotted in red.
Finally the dashed horizontal line represents an SNR equal to the 100th (10th) loudest trigger from the noise only data set.}
\label{fig:bnsfminDet}
\end{figure}

The results from these three analysis runs are shown in Fig.~\ref{fig:bnsfminDet}, where the SNR is plotted against the observed chirp mass. 
In each of the plots all the triple coincident triggers produced by \texttt{gstlal} when analysing the main data set are plotted in blue, with any of these triggers that are then matched to an injection being plotted in red and finally the triggers produced from the analysis of the noise only data set are plotted in green.

In the top plot we show the results from the 25Hz analysis where it is easy to distinguish a number of BNS signal detections from those of background events.
There is a very clear peak of triggers with low chirp masses, implying small distances, with very high SNRs.
The lower SNR events (i.e. SNR $\leq 10$) are harder to differentiate from the background events and its only by comparing them to the list of injections that we are able to identify them as true signal detections.
There is a population of higher chirp mass, high SNR triggers that have not been matched to any BNS injections and clearly are not background events.
These are in fact due to the presence of GW signals from different types of CBC within the data, in this case the population of NSBH.
This shows that the matched filtering method employed in this search is sensitive to CBC signal whose injection parameters lie outside of the search range.
Even though these are not optimal matches, as we would expect the resulting SNR to be louder than what is shown here, they are still  considered as detected. 
In these cases one would expect the recovered parameters to differ greatly from the true parameters because of the search parameter limits used when generating these template banks.
Finally we observe a large number of triggers (74,323) obtained from the noise only data set, spread across all chirp masses, with the loudest trigger having an SNR = 9.37 and the 100th loudest having an SNR = 8.566.
These are all entirely caused by the random fluctuations in the Gaussian noise data and are labelled as background events.

In the middle plot we show the results from the 10Hz analysis.
We first note here that there is a massive increase in the total number of triggers produced (341,747) which is related to the increase in the number of templates (25,252) produced for the 10Hz analysis runs compared to that of the 25Hz run (3603).
Here we clearly see the population of BNS detections that have both higher SNRs and are detectable at higher observed chirp masses.
We also note that there is a large reduction in the number of high chirp mass, high SNR unmatched detections from non-BNS signals than compared to the 25Hz analysis.
From the analysis of the noise only data set, the loudest background event has an SNR = 9.53 and the 100th loudest event has an SNR = 8.904.

In the bottom plot we show the results from the 5Hz analysis.
Again we clearly see the population of BNS signals and we also find the number of non-BNS triggers is very small.
We should also note that the number of templates has significantly increased again (87,054 templates) over that of the 10Hz analysis but we do not see as large an increase in the number of detections due to analysing only 10\% of the data.
We would expect to obtain ten times as many triggers and detections as given in Table~\ref{tab:detections}, giving an estimate of $\sim637,000$ triggers and $\sim12400$ detections from this mock data set.

Finally we highlight the loudest BNS detections in each of the analysis runs on the main data set which are produced from the same event.
Starting at 25Hz it is detected with an SNR = 98.22, at 10Hz it is detected with an SNR = 122.46 and at 5Hz it is detected with an SNR = 134.97.
This gives a clear example of how, when analysing from lower frequencies, we are able to build up more SNR for each signal which also helps us to increase the total number of detections we are able to make.

\subsection{Detection efficiency}

The detection efficiency, as a function of redshift, for a given analysis is given by

\begin{equation}
\epsilon (z) = \frac{N_\mathrm{det} (z)}{N_\mathrm{inj} (z)},
\label{eq:effCalc}
\end{equation}

\noindent where $N_\mathrm{det}$ is the number of detected injections per redshift bin, $N_\mathrm{inj}$ is the total number of injections per redshift bin and the variance is given by~\cite{cqg.25.105002.08}

\begin{equation}
\sigma_{\epsilon}^2 (z) = \frac{\epsilon (z) (1-\epsilon (z))}{N_\mathrm{inj} (z)}.
\label{eq:SigEff}
\end{equation}

\noindent In the left-hand plot of Fig.~\ref{fig:DetEfficiency} we show the smoothed detection efficiencies for each of the analysis runs carried, with the $\pm \;1\sigma$ limits contained within the shaded region.
Here we have only considered found injections that have an SNR greater than the threshold set by the 100th loudest event from the analysis of the noise only data set.
We clearly see that by lowering the cut-off frequency of the analysis we are able to increase our detection efficiency across all redhsift bins.
This can be seen clearly by the fact that the efficiency at $z=1$ doubles when going from 25Hz to 10Hz.
It is also shown that the size of the uncertainty in the 5Hz efficiency is considerably larger that for the 25Hz or 10Hz as we are only considering 10\% of the data and from Eq.~(\ref{eq:SigEff}) we see that this decreases with the inverse of the number of injections per redshift bin.

\subsection{Rate estimation}

\begin{figure*}
\includegraphics[width=0.49\textwidth]{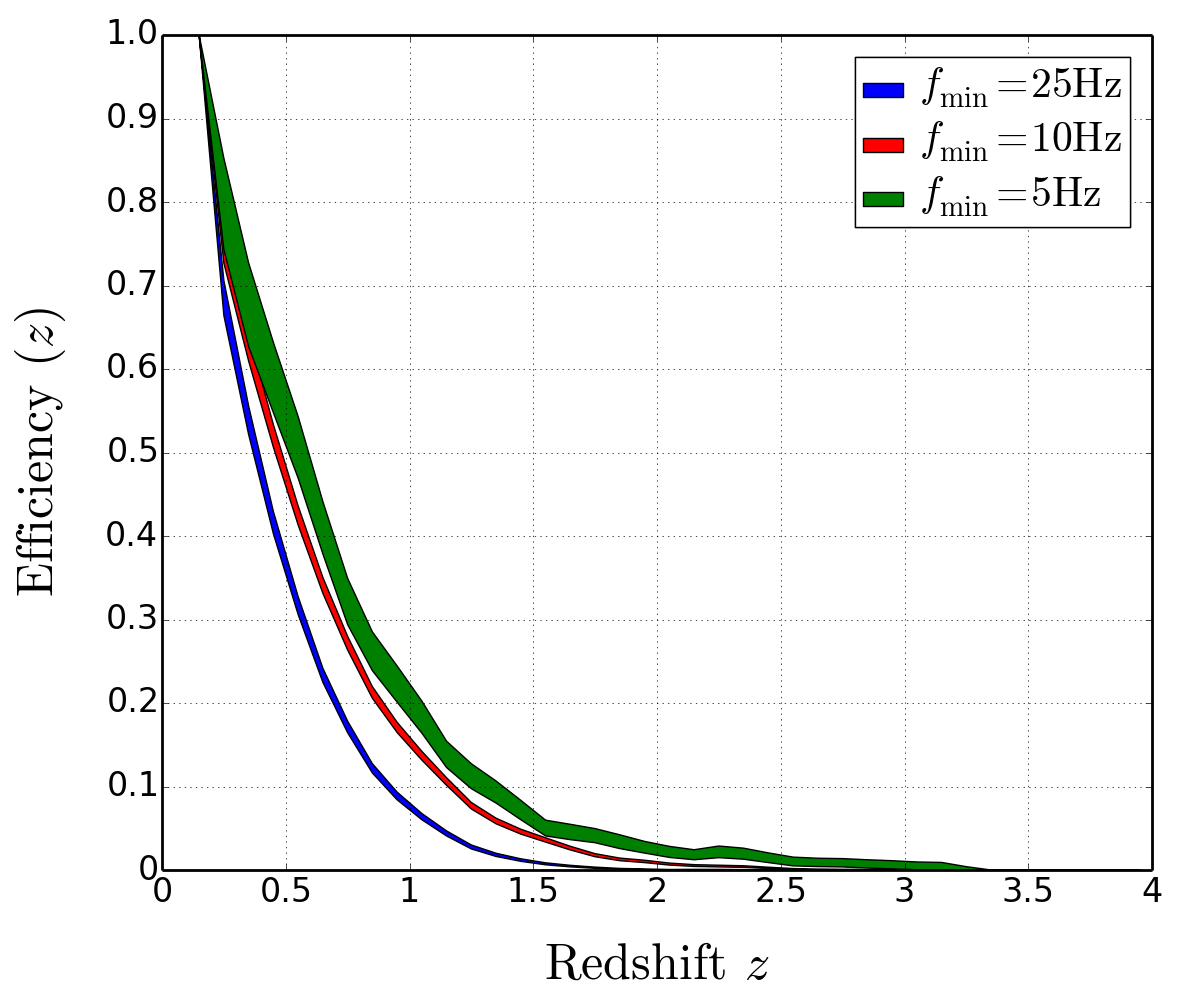}
\includegraphics[width=0.49\textwidth]{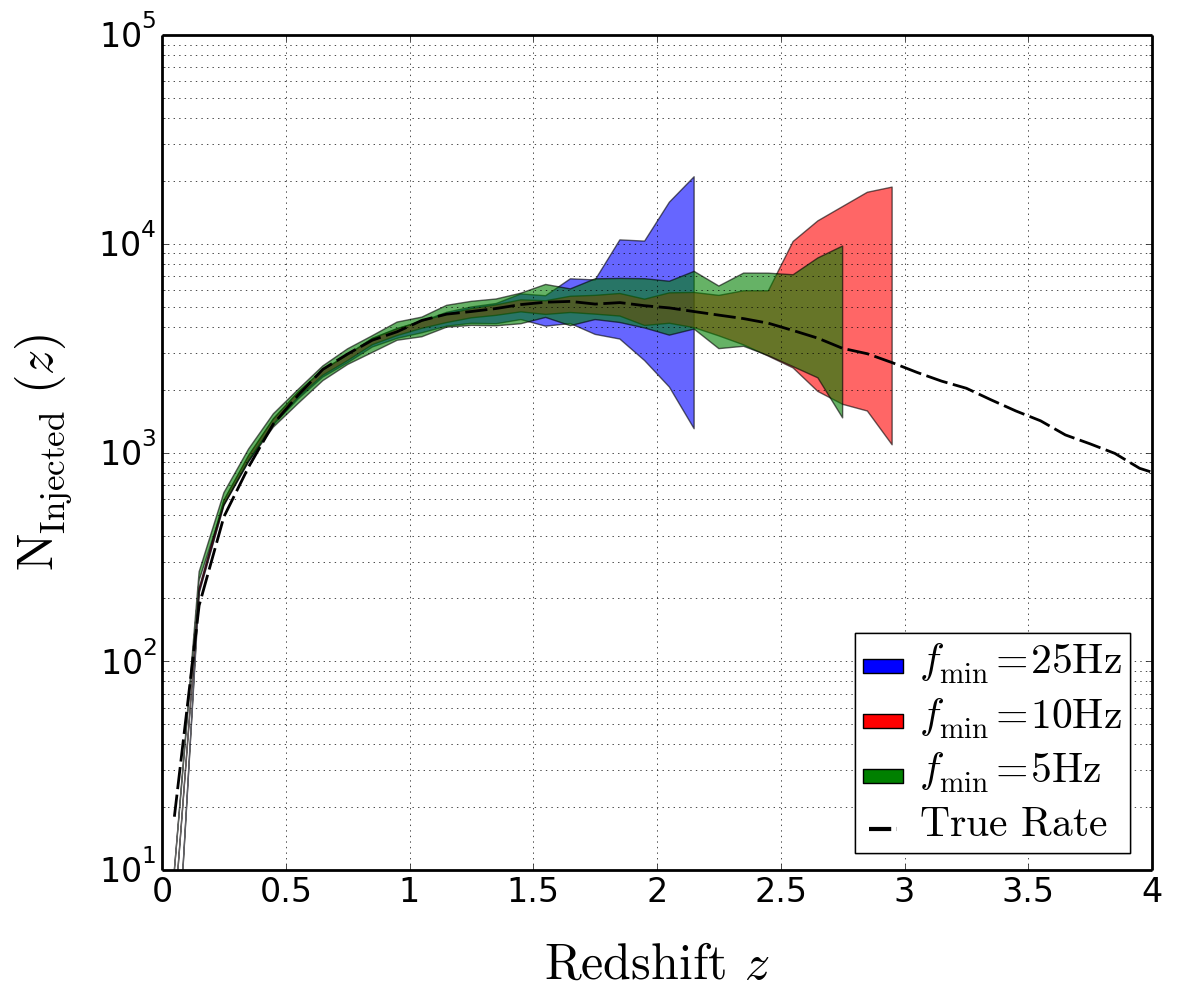}
\caption{\emph{Left}---Detection efficiency as a function of redshift for the 25Hz (blue), 10Hz (red) and 5Hz (green) analysis runs with the shaded areas representing the $\pm 1\sigma$ region, as given by Eq.~(\ref{eq:SigEff}).
\emph{Right}--- Estimation of the number of injections as a function of redshift for the 25Hz (blue), 10Hz (red) and 5Hz (green) analysis runs with the shaded areas representing the $\pm 1\sigma$ region.
The dashed black line represents the true number of injections, with redshift bins of size $\Delta z = 0.1$.}
\label{fig:DetEfficiency}
\end{figure*}

In the previous subsection we make the assumption that we know the true number and distribution of all the injections in order to calculate the efficiency. 
If we consider the case where the number of signals in the Universe is unknown, then, by rearranging Eq.~(\ref{eq:effCalc}), it is possible to make an estimate of this by consideration of the number of detections as a function of redshift\footnote{We again make the assumption that we know the true redshift of the detection. In reality we would not know the detections true redshift though it is possible to derive estimates from various methods, detailed in Section~\ref{sec:futuredev}.} along with the detection efficiency, which can be determined from MC simulations with prior knowledge of the BNS mass distribution from the second generation of detectors \cite{jpcs.484.012008.14}.
In the right-hand plot of Fig~\ref{fig:DetEfficiency} we show this estimate on the number of injections per redshift bin for each of the detection efficiencies calculated previously.
Here the errors on the size of the efficiencies have been carried through.
We clearly see that for each of the analysis runs there is a similar chance of estimating the number of events up to a redshift of $z\simeq1.5$.
Between the 25Hz (blue) and 10Hz (red) analysis runs, which were conducted on the full data set, there is a clear difference in the distance at which we are able to place an estimate on the number of injected signals, with the 25Hz extending to $z \sim 2$ and the 10Hz extending to $z \sim 3$.
This is directly related to the detection efficiency presented in the previous subsection, with the size of the estimation increasing as the efficiency goes to zero.
The 5Hz estimation appears to be larger than that of the 10Hz but this is a consequence of only analysing 10\% of the data, which results in larger uncertainties in the efficiency and a smaller maximum redshift that an estimate can be made out too.

\subsection{Impact of lower frequency cut-off on parameter estimation}

In this subsection we present the errors we obtained in the measurement of the epoch of coalescence, and binary’s chirp mass and total mass.
We first look at the absolute error in the recorded time of coalescence, given by $\Delta t_c = t_{c,\,\mathrm{obs}} - t_{c,\,\mathrm{inj}}$, followed by relative error in total mass, $M_z$, and chirp mass\footnote{We note that in the case where we know exactly the redshift of the source, the relative error in the observed masses, $M_z$ and $\mathcal{M}_z$, is mathematically identical to the relative error in the intrinsic masses, $M$ and $\mathcal{M}$.}, $\mathcal{M}_z$.
Table~\ref{tab:errors} lists the values of the mean and standard deviation for all the errors shown in this section.

\subsubsection{Coalescence time}

\begin{figure}
\includegraphics[width=0.5\textwidth]{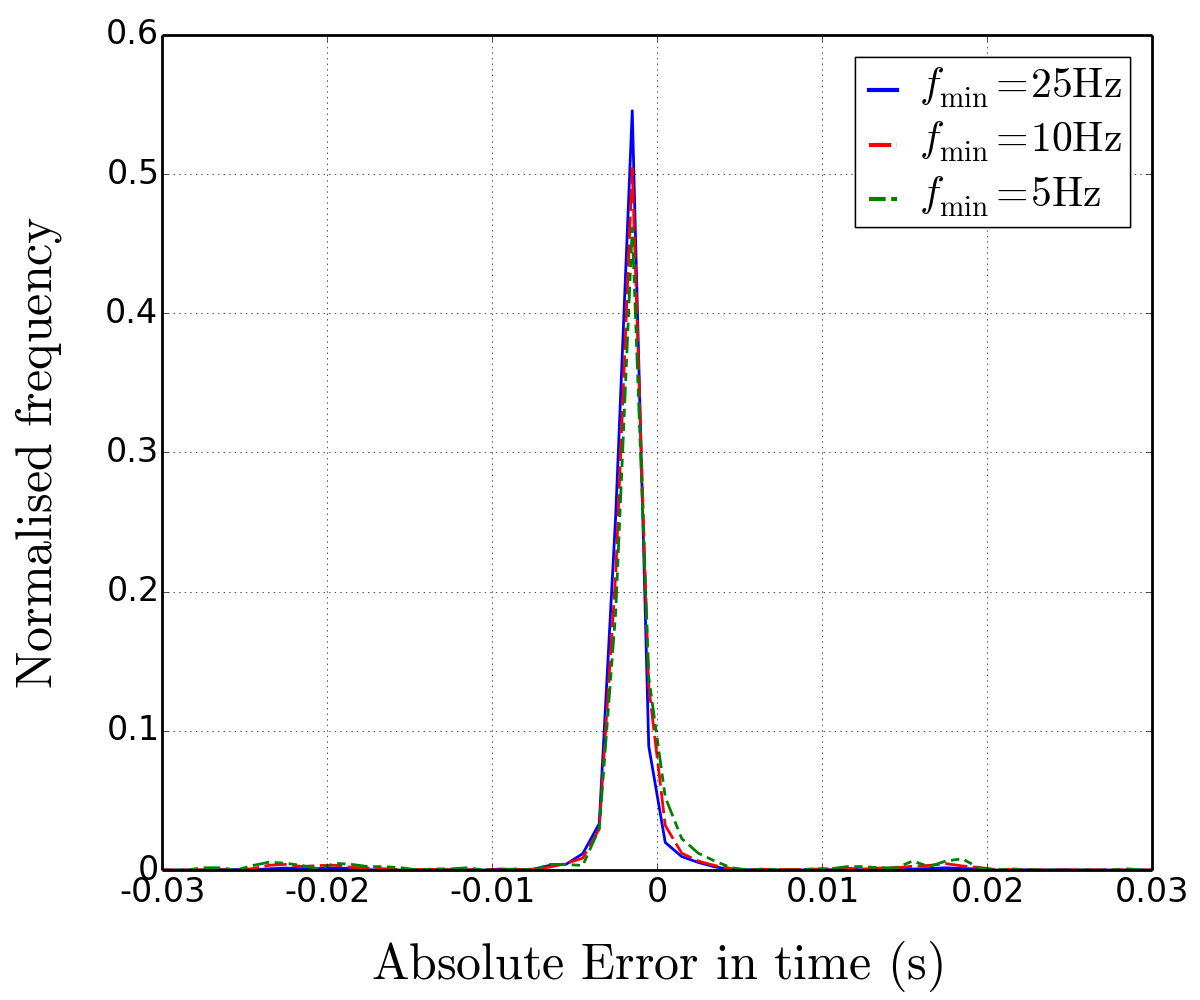}
\caption{Normalised distribution of absolute error in recovered coalescence time for all matched detections given by \texttt{gstlal} for Search 2 at 25Hz (solid blue), Search 3 at 10Hz (dashed red), and Search 4 at 5Hz (dot-dashed green), using time bins of size $\Delta t$ = 1 ms, where different low frequency cut-offs were used.}
\label{fig:abstcErr}
\end{figure}

In this first MDSC, when matching triggers to injections, we considered a time window of $\pm$30ms while in this investigation, as stated above, we have increased this to $\pm$100ms.
In Fig.~\ref{fig:abstcErr} we show a normalised plot of absolute error in measured coalescence time, $t_c$, of all the detections made when investigating the low frequency cut-off.
We find that for all three BNS runs there is a constant bias of a few ms but nearly all detections are constrained very well to within $\pm$10ms.
This is due to the fact that both the injected waveform and the waveform used to search the data end at the same point, the $f_\mathrm{lsco}$.
So the $\pm$30 ms window considered for the first MDSC is suitable when considering BNS signals.

\subsubsection{Masses}
\label{sec:fminPE}

\begin{figure*}
\includegraphics[width=0.49\textwidth]{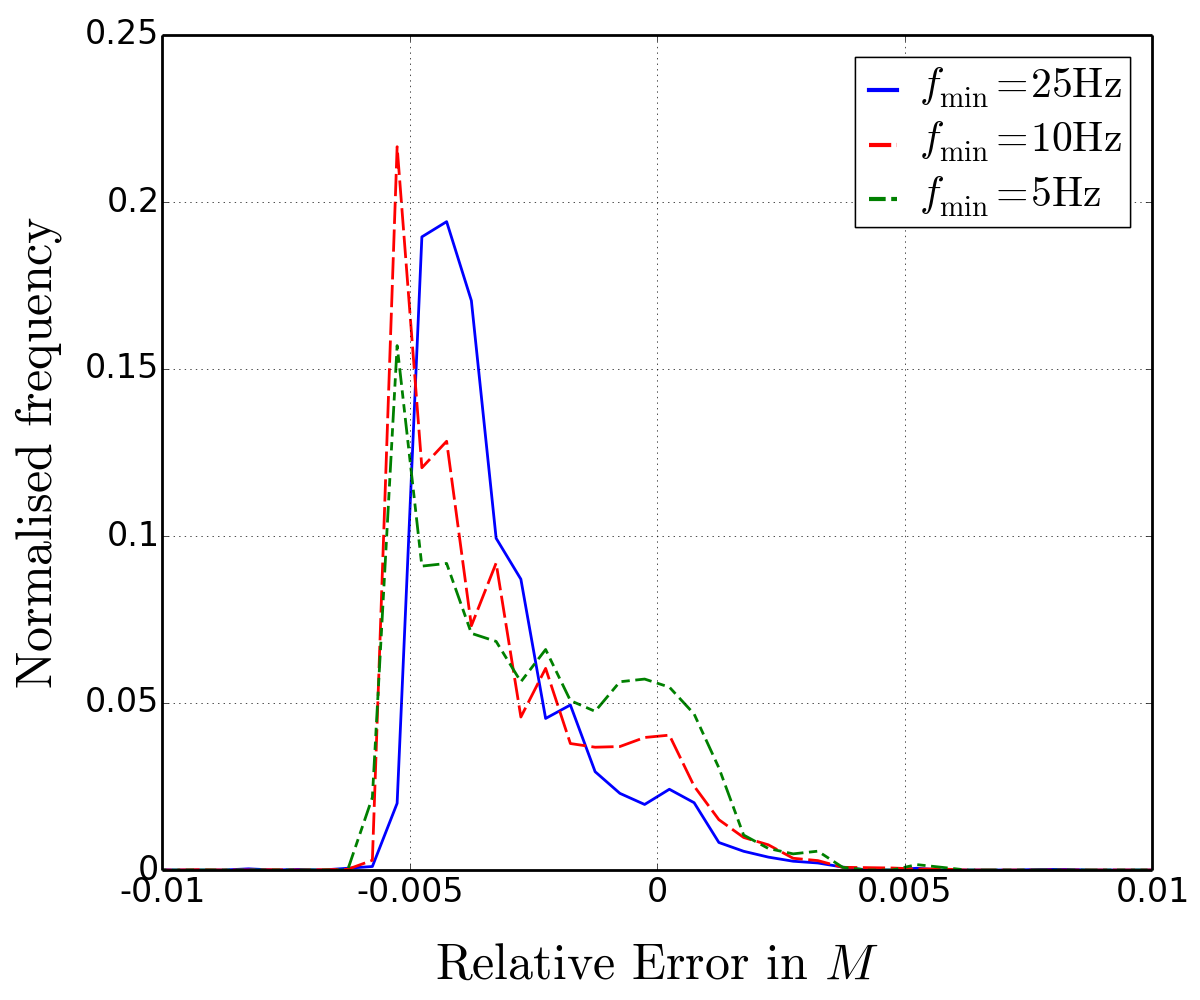}
\includegraphics[width=0.49\textwidth]{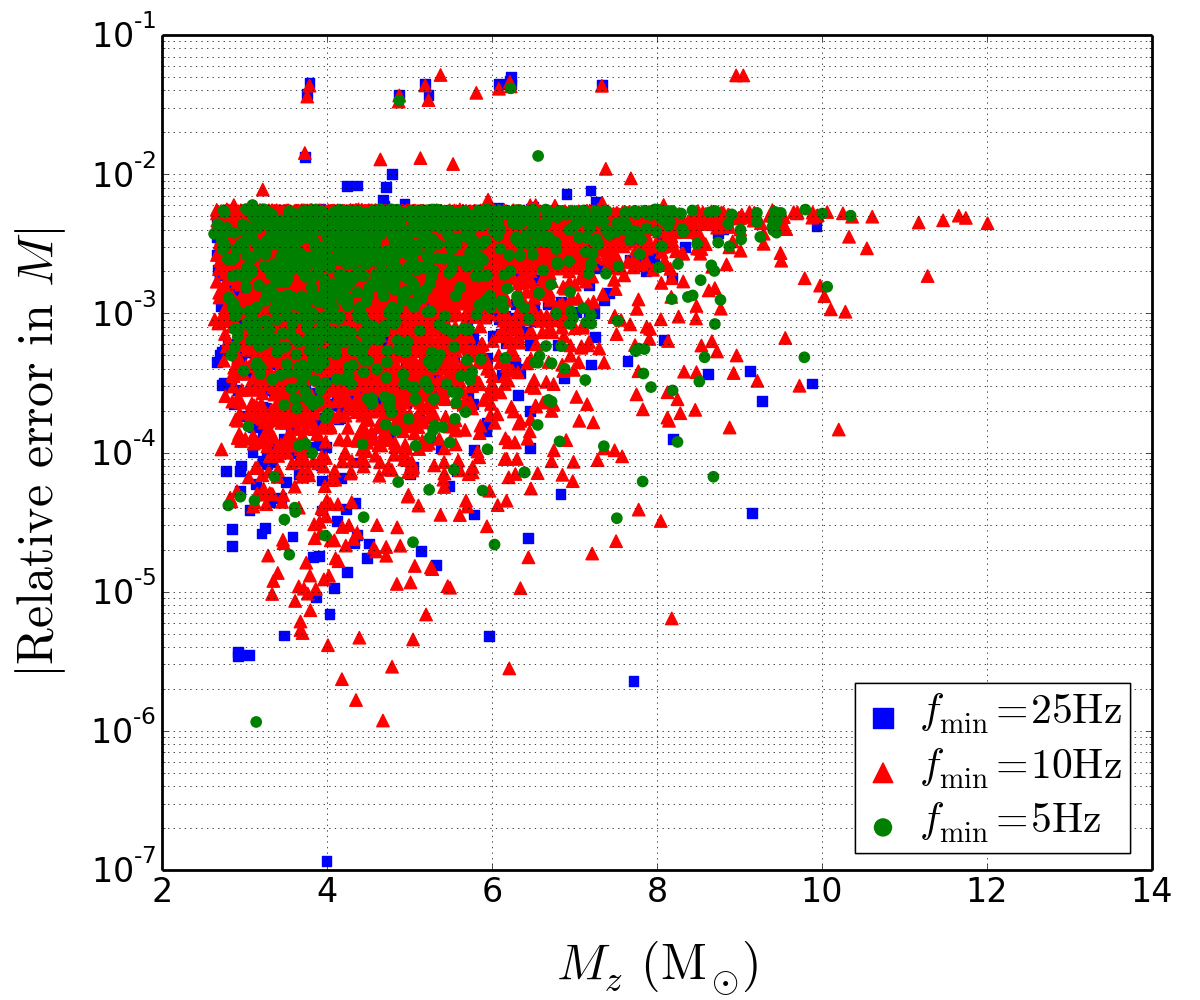} \\
\includegraphics[width=0.49\textwidth]{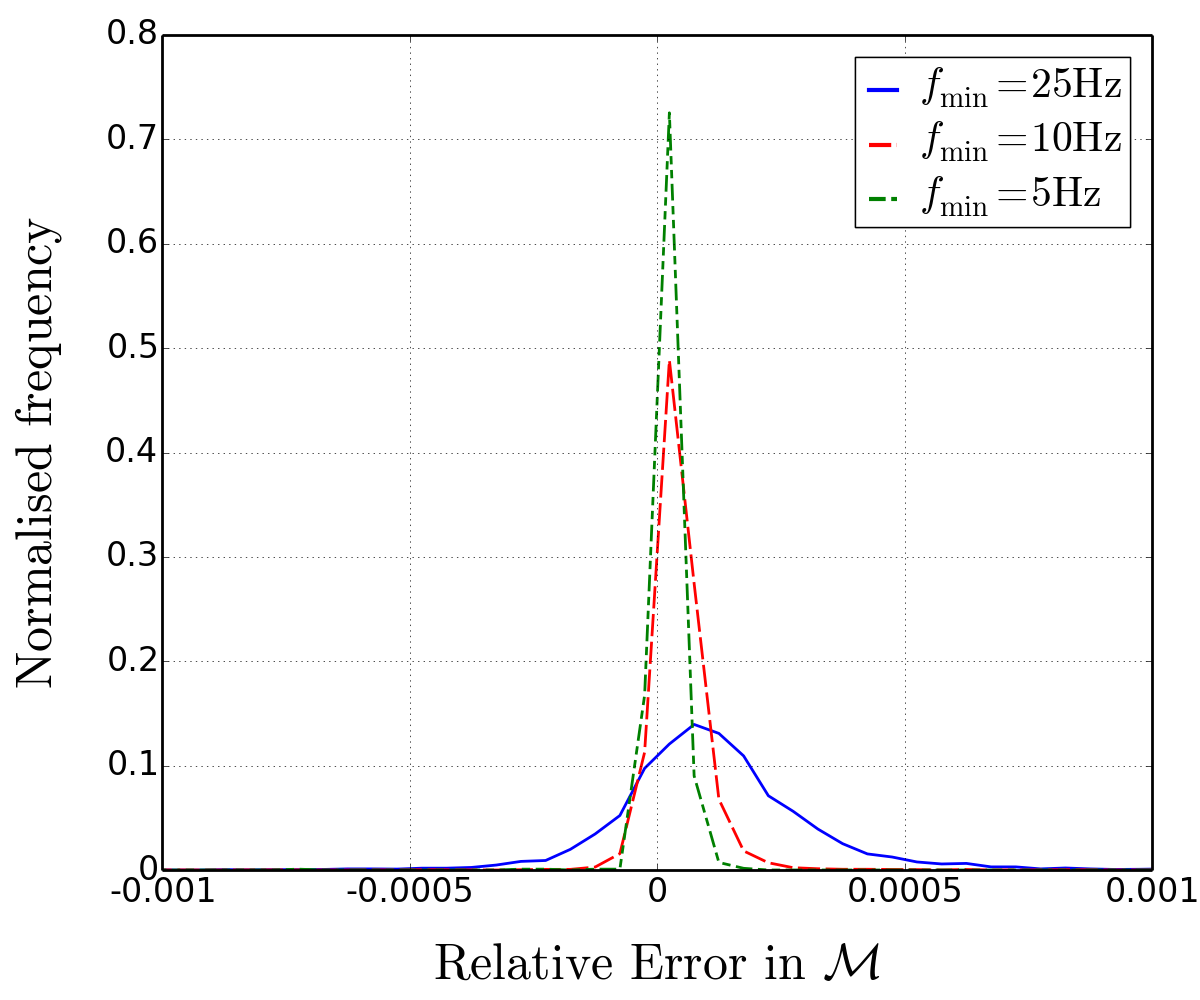}
\includegraphics[width=0.49\textwidth]{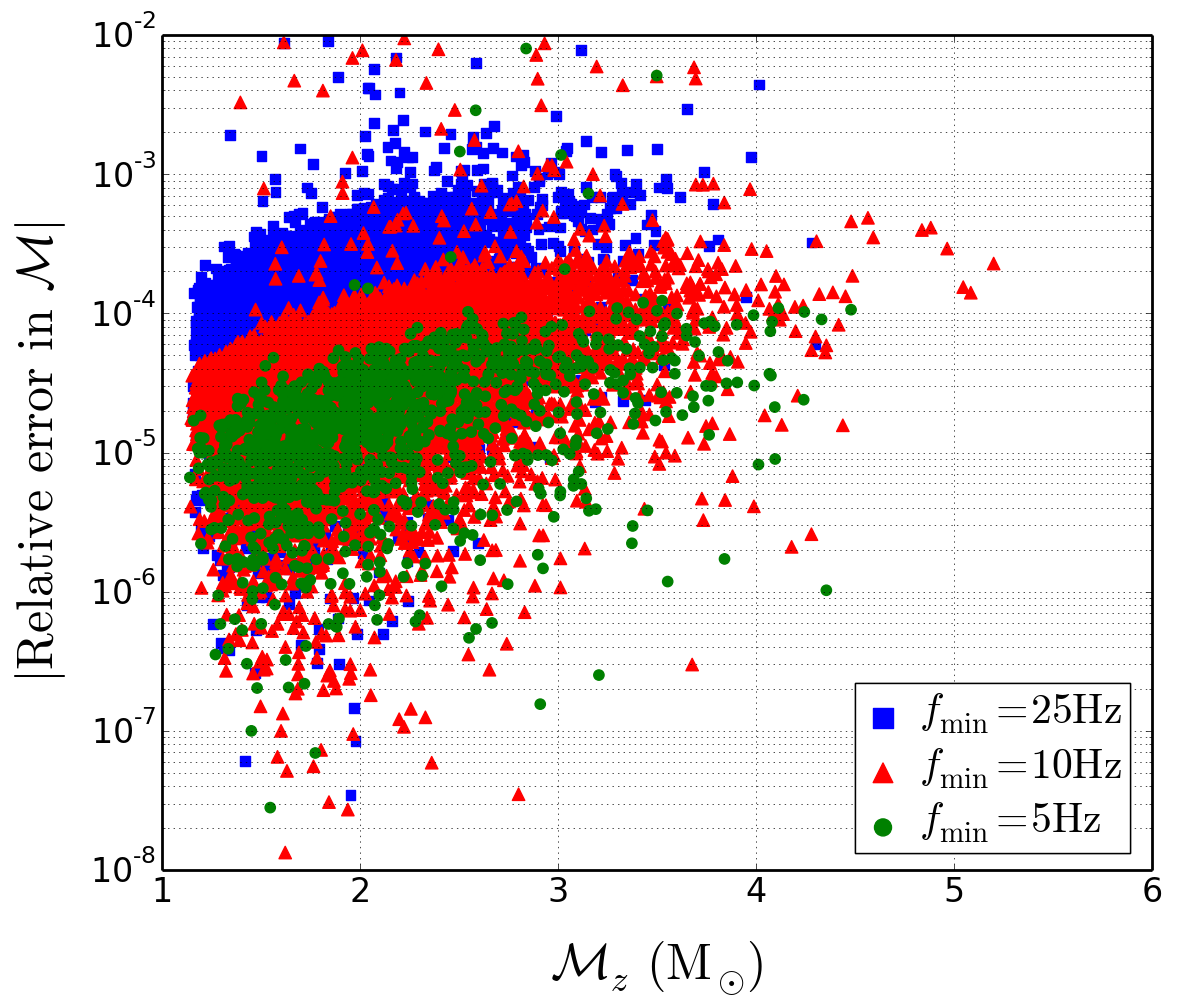}
\caption{\emph{Top left}--- Normalised distribution of relative error in recovered total mass for Search 2 at 25Hz (solid blue), Search 3 at 10Hz (dashed red), and Search 4 at 5Hz (dot-dashed green), using mass error bins of size $\Delta M = 5\times10^{-4}$.
\emph{Top right}--- Scatter plot of relative error in total mass as a function of the observed total mass for Search 2 at 25Hz (blue $\blacksquare$), Search 3 at 10Hz (red $\blacktriangle$), and Search 4 at 5Hz (green $\bullet$).
\emph{Bottom left}--- Normalised distribution of relative error in recovered chirp mass for Search 2 at 25Hz (solid blue), Search 3 at 10Hz (dashed red), and Search 4 at 5Hz (dot-dashed green), using mass error bins of size $\Delta \mathcal{M} = 5\times10^{-5}$.
\emph{Bottom right}--- Scatter plot of relative error in chirp mass as a function of the observed chirp mass for Search 2 at 25Hz (blue $\blacksquare$), Search 3 at 10Hz (red $\blacktriangle$), and Search 4 at 5Hz (green $\bullet$).}
\label{fig:fminError}
\end{figure*}

We now look at the errors in the measurements of the mass parameters. 
In Fig.~\ref{fig:fminError} we show the impact of lowering the minimum search frequency.

In the top left-hand plot of Fig.~\ref{fig:fminError} we show a normalised distribution of the relative error in measured total mass with the results from the 25Hz analysis shown in blue, the results from the 10Hz analysis shown in red and the results from the 5Hz analysis shown in green.
We first note that the error has decreased by an order of magnitude when compared to the results from the first MDSC (see Fig.~7 of~\cite{prd.86.122001.12}).
Also there is a constant systematic bias to generally underestimate the total mass for all three analysis runs, with a sudden drop off below 0.5\%.
The number of events where the total mass is underestimated does decrease as the cut-off frequency for the analysis is lowered but this is only a small proportion.
This bias was not observed in either the first MDSC or in any of our initial analysis runs where, in both cases, the component masses, $m_1$ and $m_2$, were selected from the same distribution, which is not the case for this main mock data set.

In the top right-hand figure we plot the relative error in total mass against the observed total mass with the results from the 25Hz analysis shown in blue, the results from the 10Hz analysis shown in red and the results from the 5Hz analysis shown in green. 
We clearly see that sharp cut-off at the 0.5\% shown in the previous plot.
We also see that at lower observed masses, which correspond to closer distances, the spread of error measurements covers a range of values.
At higher masses this distribution decreases leaving only the larger error measurements.
This agrees with what we would expect, that our error measurements increase with distance.

In the bottom left-hand plot we show a normalised distribution of the relative error in measured chirp mass with the results from the 25Hz analysis shown in blue, the results from the 10Hz analysis shown in red and the results from the 5Hz analysis shown in green.
We first note that the scale of the size of the distribution of the error has also decreased by a factor of $\sim10$ when compared to the results from the first MDSC.
Here we clearly see that as we decrease the cut-off frequency for the analysis we obtain a smaller distribution of the error of the chirp mass measurement.
We can also see from Table~\ref{tab:errors} that the deviation of the mean of the distribution from zero goes from 0.01\% at 25Hz to 0.001\% at 5Hz which shows that we are able to recover the chirp mass to a very high degree of accuracy in this part of the analysis.

In the bottom right-hand figure we plot the relative error in chirp mass against the observed chirp mass with the results from the 25Hz analysis shown in blue, the results from the 10Hz analysis shown in red and the results from the 5Hz analysis shown in green.
Here we clearly see that by decreasing the cut-off frequency we are able to better measure the chirp mass but also that the measured error on the chirp mass is related to the distance to the source.

\begin{table*}
\caption{\label{tab:errors} Table showing the mean and standard deviations of the error in the measurements of injection parameters.
The first column indicates which search it is.
The second column gives the mean and standard deviations of the absolute error in measured coalescence time in milliseconds.
The third column gives the mean and standard deviations of the relative error in the measurement of the total mass.
The third column gives the mean and standard deviations of the relative error in the measurement of the chirp mass.}
\begin{center}
\begin{tabular}{ l c c c}
\hline \hline
Search    & $\Delta t_c$ (ms)  & Relative error $M$ & Relative error $\mathcal{M}$ \\ \hline
2 (25Hz) \,\,\, & \,\,\,  -1.694   $\pm$ 3.314  \,\,\,  & \,\,\, -3.301  $\times 10^{-3} \pm$ 2.353   $\times 10^{-3}$ \,\,\, & \,\,\,  0.115 $\times 10^{-3} \pm$ 0.369 $\times 10^{-3}$ \,\,\, \\
3 (10Hz) &  -1.541   $\pm$ 5.307   & -3.213   $\times 10^{-3} \pm$ 2.550   $\times 10^{-3}$ &   0.044 $\times 10^{-3} \pm$ 0.286 $\times 10^{-3}$ \\
4 (5Hz)   &  -1.572   $\pm$ 5.856   & -2.674   $\times 10^{-3} \pm$ 2.665   $\times 10^{-3}$ &   0.012 $\times 10^{-3} \pm$ 0.289 $\times 10^{-3}$ \\
\hline \hline
\end{tabular}
\end{center}
\end{table*}


\section{Future Development}
\label{sec:futuredev}

Future MDSCs should aim to address increasing complexity of binary waveform models, improved detector noise models, simulating EM counterpart scenarios, and including other third generation detectors. 
There are still other GW sources that we can consider, consider, such as continuous waves~\cite{apj.785.119.14} from rapidly rotating galactic neutron stars~\cite{prd.88.102002.13,prd.90.062010.14}.
The inclusion of one or more SGWBs of cosmological origins~\cite{jcap.6.27.12}, such as phase transitions~\cite{prd.77.124015.08,prd.79.083519.09,jcap.12.024.09}, cosmic (super) strings~\cite{prd.71.063510.05,prl.98.111101.07,prd.81.104028.10,prd.85.066001.12,prl.112.131101.13} or pre Big Bang models~\cite{app.1.317.93,prd.55.3330.97,prd.82.083518.10}, would allow us to test whether we can distinguish between cosmological background and astrophysical backgrounds~\cite{prd.85.104024.12}.
The waveform models that we choose to inject should also include additional features such as spin \cite{prd.72.084027.05,prd.67.104025.06,prd.74.029902.06} and tidal affects \cite{prd.77.021502.08,prd.79.124033.09,prd.81.123016.10,prd.84.104017.11}, for BNS and NSBH, spin and precession \cite{prl.113.151101.14,prd.91.024043.15}, for BBH and IMBH, and use a larger range of burst signal models.
The inspiral waveforms should be generated down to even lower frequencies, such as 3Hz or 1Hz, to investigate if it is possible to push the low frequency cut-off used for the matched filtering past the 5Hz used here.
At this frequency the low mass waveforms will be of the order of $\sim$ hours to days long.
These would allow for investigations into areas such as rate estimation, both the SFR and coalescence rate for various sources, measurement of the mass functions for NSBH and BBH, testing of general relativity, cosmological measurements, investigating different cosmological and astrophysical models and testing alternate theories of gravity.

When generating the data we should also include the two LIGO detectors with the use of the LIGO 3 Strawman PSD~\cite{StrawmanRed}.
A smaller second data set should also be constructed with the use of re-coloured aLIGO noise (which we would expect to have at that point) into which we inject coherent signals.
This will allow to study the behaviour of the null stream in the non-Gaussian case.

It is impossible to obtain a redshift measurement directly from a detection of a GW but it is possible to infer one through the use of an electromagnetic counterpart such as a  sGRBs~\cite{apj.725.496.10} or from an existing galaxy catalogue~\cite{prd.86.043011.12}, or consideration of either the neutron star mass function~\cite{prd.86.023502.12}, or EOS~\cite{prl.108.091101.12}.
None of these methods have yet been applied within an MDSC, but some of them, such as using sGRBs, the neutron star mass function, or EOS, can easily be included within a future MDSC.


\section{Conclusion}
\label{sec:conclusion}

In this investigation we have described the generation and analysis of the data for the second Einstein Telescope mock data and science challenge with a focus on binary neutron stars.
This data consisted of Gaussian noise, fitted to the expected ET-D sensitivity noise curve, into which a large number of GW signals from multiple sources are injected.
The analysis was conducted with a new matched filtering pipeline that is able to analyse signals down to lower frequencies than has been considered before.
Our motivation for this MDSC is to continue to explore the science potential of ET, increasing the complexity of the data analysis and science that is conducted with it.

The analysis used in this investigation has far surpassed that carried out in the first MDSC.
One of the main goals for this investigation was to show that it is possible to analyse gravitational-wave inspiral signals down to a frequency of 5Hz.
Starting at this frequency the lowest mass BNS systems being considered here take over two hours to coalesce.
We have shown that, while being very computationally intensive/expensive, it is still possible to analyse data down to this frequency.
If we consider that in the few years since the first MDSC we have been able to push the limit of the analysis comfortably from 25Hz to 10Hz and proven that 5Hz is achievable, we would like to think that in the next decade when the Einstein Telescope is hoped to be built, given Moore's law, it should be possible to push GW analysis to even lower frequency limits.

In the analysis at lower frequencies we have also shown the improvement we obtain in both detection efficiency and our ability to recover the injection parameters.
By searching for signals with lower frequencies we are able build up more SNR which allows many more signals to become detectable as well as making the already detectable signals louder.
The longer template waveforms also allow us to better match up with the GW signals, giving us better accuracy in the measurements of the parameters.

It has also been shown that analysing data at lower frequencies results in a higher rate of background detections being made with larger SNRs.
Here we have just considered using an SNR threshold values that is equal to the 100th (10th) loudest background event from the analysis of the noise only data set, to reduce the number of background events but this has the drawback of reducing the number of true detections that are made as well.
In the future it is hoped that a method will be developed that implements the null stream to reject background events, thus lowering the false alarm probability, allowing for a smaller SNR threshold to be used.

We have also shown the difference in detection efficiencies obtained when using lower cut-off frequencies.
From these a proof of concept method has been shown where we attempt to estimate the number of injected signals as a function of redshift.
This is a very basic method that makes several assumptions, mainly that we know the true redshifts of the detected signals.
More work is required to further develop this method so that it is able to account for different parameters as well as a distribution on the redshift from the detections.

Finally we have also shown that our ability to measure mass parameters improved by an order of magnitude over that of the first MDSC in the case of BNS as a result of using a 5 Hz lower frequency cut-off instead of 25 Hz.
We are able to recover the observed total mass to within 0.5\% and the observed chirp mass to within 0.05\%.

This work will now continue, were we investigate the parameter estimation for a small subset of the BNS detections.


\section{Acknowledgements}

We thank Bruce Allen and the Albert Einstein Institute in Hannover, supported by the Max-Planck-Gesellschaft, for use of the Atlas high-performance computing cluster in the data generation and analysis, and Carsten Aulbert for technical advice and assistance.
DM acknowledges the PhD financial support from the Observatoire de la C\^{o}te d’Azur and the PACA region and would also like to thank Cardiff university for funding under which part of this work was conducted.
CH is supported by NSF grant PHY-1454389.
BSS acknowledges the support of the LIGO Visitor Program through the National Science Foundation award PHY-0757058, Max-Planck Institute of Gravitational Physics, Potsdam, Germany, and STFC grant ST/J000345/1.


\bibliography{bibliography}

\begin{thebibliography}{79}
\expandafter\ifx\csname natexlab\endcsname\relax\def\natexlab#1{#1}\fi
\expandafter\ifx\csname bibnamefont\endcsname\relax
  \def\bibnamefont#1{#1}\fi
\expandafter\ifx\csname bibfnamefont\endcsname\relax
  \def\bibfnamefont#1{#1}\fi
\expandafter\ifx\csname citenamefont\endcsname\relax
  \def\citenamefont#1{#1}\fi
\expandafter\ifx\csname url\endcsname\relax
  \def\url#1{\texttt{#1}}\fi
\expandafter\ifx\csname urlprefix\endcsname\relax\def\urlprefix{URL }\fi
\providecommand{\bibinfo}[2]{#2}
\providecommand{\eprint}[2][]{\url{#2}}

\bibitem[{\citenamefont{Aasi et~al.}(2015)}]{cqg.32.074001.15}
\bibinfo{author}{\bibfnamefont{J.}~\bibnamefont{Aasi}} \bibnamefont{et~al.}
  (\bibinfo{collaboration}{LIGO Scientific Collaboration}),
  \bibinfo{journal}{Class. Quantum Grav.} \textbf{\bibinfo{volume}{32}},
  \bibinfo{pages}{074001} (\bibinfo{year}{2015}).

\bibitem[{\citenamefont{Acernese et~al.}(2015)}]{cqg.32.024001.15}
\bibinfo{author}{\bibfnamefont{F.}~\bibnamefont{Acernese}} \bibnamefont{et~al.}
  (\bibinfo{collaboration}{Virgo Collaboration}), \bibinfo{journal}{Class.
  Quantum Grav.} \textbf{\bibinfo{volume}{32}}, \bibinfo{pages}{024001}
  (\bibinfo{year}{2015}).

\bibitem[{\citenamefont{Abbott et~al.}(2009{\natexlab{a}})}]{rpp.72.076901.09}
\bibinfo{author}{\bibfnamefont{B.~P.} \bibnamefont{Abbott}}
  \bibnamefont{et~al.} (\bibinfo{collaboration}{LIGO Scientific
  Collaboration}), \bibinfo{journal}{Rep. Prog. Phys.}
  \textbf{\bibinfo{volume}{72}}, \bibinfo{pages}{076901}
  (\bibinfo{year}{2009}{\natexlab{a}}).

\bibitem[{\citenamefont{Acernese et~al.}(2005)}]{aip.794.307.05}
\bibinfo{author}{\bibfnamefont{F.}~\bibnamefont{Acernese}} \bibnamefont{et~al.}
  (\bibinfo{collaboration}{Virgo Collaboration}), \bibinfo{journal}{AIP Conf.
  Proc.} \textbf{\bibinfo{volume}{794}}, \bibinfo{pages}{307}
  (\bibinfo{year}{2005}).

\bibitem[{\citenamefont{Punturo et~al.}(2010)}]{cqg.27.194002.10}
\bibinfo{author}{\bibfnamefont{M.}~\bibnamefont{Punturo}} \bibnamefont{et~al.},
  \bibinfo{journal}{Class. Quantum Grav.} \textbf{\bibinfo{volume}{27}},
  \bibinfo{pages}{194002} (\bibinfo{year}{2010}).

\bibitem[{\citenamefont{Buonanno et~al.}(2005)\citenamefont{Buonanno, Sigl,
  Raffelt, Janka, and {M{\"u}ller}}}]{prd.72.084001.05}
\bibinfo{author}{\bibfnamefont{A.}~\bibnamefont{Buonanno}},
  \bibinfo{author}{\bibfnamefont{G.}~\bibnamefont{Sigl}},
  \bibinfo{author}{\bibfnamefont{G.~G.} \bibnamefont{Raffelt}},
  \bibinfo{author}{\bibfnamefont{H.~T.} \bibnamefont{Janka}}, \bibnamefont{and}
  \bibinfo{author}{\bibfnamefont{E.}~\bibnamefont{{M{\"u}ller}}},
  \bibinfo{journal}{Phys. Rev. D} \textbf{\bibinfo{volume}{72}},
  \bibinfo{pages}{084001} (\bibinfo{year}{2005}).

\bibitem[{\citenamefont{Sandick et~al.}(2006)\citenamefont{Sandick, Olive,
  Daigne, and Vangioni}}]{prd.73.104024.06}
\bibinfo{author}{\bibfnamefont{P.}~\bibnamefont{Sandick}},
  \bibinfo{author}{\bibfnamefont{K.~A.} \bibnamefont{Olive}},
  \bibinfo{author}{\bibfnamefont{F.}~\bibnamefont{Daigne}}, \bibnamefont{and}
  \bibinfo{author}{\bibfnamefont{E.}~\bibnamefont{Vangioni}},
  \bibinfo{journal}{Phys. Rev. D} \textbf{\bibinfo{volume}{73}},
  \bibinfo{pages}{104024} (\bibinfo{year}{2006}).

\bibitem[{\citenamefont{Marassi et~al.}(2009)\citenamefont{Marassi, Schneider,
  and Ferrari}}]{mnras.398.293.09}
\bibinfo{author}{\bibfnamefont{S.}~\bibnamefont{Marassi}},
  \bibinfo{author}{\bibfnamefont{R.}~\bibnamefont{Schneider}},
  \bibnamefont{and} \bibinfo{author}{\bibfnamefont{V.}~\bibnamefont{Ferrari}},
  \bibinfo{journal}{MNRAS} \textbf{\bibinfo{volume}{398}}, \bibinfo{pages}{293}
  (\bibinfo{year}{2009}).

\bibitem[{\citenamefont{Zhu et~al.}(2010)\citenamefont{Zhu, Howell, and
  Blair}}]{mnrasl.409.L132.10}
\bibinfo{author}{\bibfnamefont{X.~J.} \bibnamefont{Zhu}},
  \bibinfo{author}{\bibfnamefont{E.}~\bibnamefont{Howell}}, \bibnamefont{and}
  \bibinfo{author}{\bibfnamefont{D.}~\bibnamefont{Blair}},
  \bibinfo{journal}{MNRASL} \textbf{\bibinfo{volume}{409}},
  \bibinfo{pages}{L132} (\bibinfo{year}{2010}).

\bibitem[{\citenamefont{Regimbau and
  de~Freitas~Pacheco}(2001)}]{aap.376.381.01}
\bibinfo{author}{\bibfnamefont{T.}~\bibnamefont{Regimbau}} \bibnamefont{and}
  \bibinfo{author}{\bibfnamefont{J.~A.} \bibnamefont{de~Freitas~Pacheco}},
  \bibinfo{journal}{Astron. \& Astrophys.} \textbf{\bibinfo{volume}{376}},
  \bibinfo{pages}{381} (\bibinfo{year}{2001}).

\bibitem[{\citenamefont{Rosado}(2012)}]{prd.86.104007.12}
\bibinfo{author}{\bibfnamefont{P.~A.} \bibnamefont{Rosado}},
  \bibinfo{journal}{Phys. Rev. D} \textbf{\bibinfo{volume}{86}},
  \bibinfo{pages}{104007} (\bibinfo{year}{2012}).

\bibitem[{\citenamefont{Rosado}(2011)}]{prd.84.084004.11}
\bibinfo{author}{\bibfnamefont{P.~A.} \bibnamefont{Rosado}},
  \bibinfo{journal}{Phys. Rev. D} \textbf{\bibinfo{volume}{84}},
  \bibinfo{pages}{084004} (\bibinfo{year}{2011}).

\bibitem[{\citenamefont{Marassi et~al.}(2011)\citenamefont{Marassi, Schneider,
  Corvino, Ferrari, and Zwart}}]{prd.84.124037.11}
\bibinfo{author}{\bibfnamefont{S.}~\bibnamefont{Marassi}},
  \bibinfo{author}{\bibfnamefont{R.}~\bibnamefont{Schneider}},
  \bibinfo{author}{\bibfnamefont{G.}~\bibnamefont{Corvino}},
  \bibinfo{author}{\bibfnamefont{V.}~\bibnamefont{Ferrari}}, \bibnamefont{and}
  \bibinfo{author}{\bibfnamefont{S.~P.} \bibnamefont{Zwart}},
  \bibinfo{journal}{Phys. Rev. D} \textbf{\bibinfo{volume}{84}},
  \bibinfo{pages}{124037} (\bibinfo{year}{2011}).

\bibitem[{\citenamefont{Sathyaprakash et~al.}(2012)}]{cqg.29.124013.12}
\bibinfo{author}{\bibfnamefont{B.~S.} \bibnamefont{Sathyaprakash}}
  \bibnamefont{et~al.}, \bibinfo{journal}{Class. Quantum Grav.}
  \textbf{\bibinfo{volume}{29}}, \bibinfo{pages}{124013}
  (\bibinfo{year}{2012}).

\bibitem[{\citenamefont{Regimbau
  et~al.}(2012{\natexlab{a}})\citenamefont{Regimbau, Dent, Del~Pozzo,
  Giampanis, Li, Robinson, Van Den~Broeck, Meacher, Rodriguez, Sathyaprakash
  et~al.}}]{prd.86.122001.12}
\bibinfo{author}{\bibfnamefont{T.}~\bibnamefont{Regimbau}},
  \bibinfo{author}{\bibfnamefont{T.}~\bibnamefont{Dent}},
  \bibinfo{author}{\bibfnamefont{W.}~\bibnamefont{Del~Pozzo}},
  \bibinfo{author}{\bibfnamefont{S.}~\bibnamefont{Giampanis}},
  \bibinfo{author}{\bibfnamefont{T.~G.~F.} \bibnamefont{Li}},
  \bibinfo{author}{\bibfnamefont{C.}~\bibnamefont{Robinson}},
  \bibinfo{author}{\bibfnamefont{C.}~\bibnamefont{Van Den~Broeck}},
  \bibinfo{author}{\bibfnamefont{D.}~\bibnamefont{Meacher}},
  \bibinfo{author}{\bibfnamefont{C.}~\bibnamefont{Rodriguez}},
  \bibinfo{author}{\bibfnamefont{B.~S.} \bibnamefont{Sathyaprakash}},
  \bibnamefont{et~al.}, \bibinfo{journal}{Phys. Rev. D}
  \textbf{\bibinfo{volume}{86}}, \bibinfo{pages}{122001}
  (\bibinfo{year}{2012}{\natexlab{a}}).

\bibitem[{\citenamefont{Abbott et~al.}(2009{\natexlab{b}})}]{prd.79.122001.09}
\bibinfo{author}{\bibfnamefont{B.~P.} \bibnamefont{Abbott}}
  \bibnamefont{et~al.} (\bibinfo{collaboration}{LIGO Scientific
  Collaboration}), \bibinfo{journal}{Phys. Rev. D}
  \textbf{\bibinfo{volume}{79}}, \bibinfo{pages}{122001}
  (\bibinfo{year}{2009}{\natexlab{b}}).

\bibitem[{\citenamefont{Abbott et~al.}(2009{\natexlab{c}})}]{prd.80.047101.09}
\bibinfo{author}{\bibfnamefont{B.~P.} \bibnamefont{Abbott}}
  \bibnamefont{et~al.} (\bibinfo{collaboration}{LIGO Scientific
  Collaboration}), \bibinfo{journal}{Phys. Rev. D}
  \textbf{\bibinfo{volume}{80}}, \bibinfo{pages}{047101}
  (\bibinfo{year}{2009}{\natexlab{c}}).

\bibitem[{\citenamefont{Abadie et~al.}(2010{\natexlab{a}})}]{prd.82.102001.10}
\bibinfo{author}{\bibfnamefont{J.}~\bibnamefont{Abadie}} \bibnamefont{et~al.}
  (\bibinfo{collaboration}{LIGO Scientific Collaboration and Virgo
  Collaboration}), \bibinfo{journal}{Phys. Rev. D}
  \textbf{\bibinfo{volume}{82}}, \bibinfo{pages}{102001}
  (\bibinfo{year}{2010}{\natexlab{a}}).

\bibitem[{\citenamefont{Abadie et~al.}(2012)}]{prd.85.082002.12}
\bibinfo{author}{\bibfnamefont{J.}~\bibnamefont{Abadie}} \bibnamefont{et~al.}
  (\bibinfo{collaboration}{LIGO Scientific Collaboration and Virgo
  Collaboration}), \bibinfo{journal}{Phys. Rev. D}
  \textbf{\bibinfo{volume}{85}}, \bibinfo{pages}{082002}
  (\bibinfo{year}{2012}).

\bibitem[{\citenamefont{Allen and Romano}(1999)}]{prd.59.102001.99}
\bibinfo{author}{\bibfnamefont{B.}~\bibnamefont{Allen}} \bibnamefont{and}
  \bibinfo{author}{\bibfnamefont{J.~D.} \bibnamefont{Romano}},
  \bibinfo{journal}{Phys. Rev. D} \textbf{\bibinfo{volume}{59}},
  \bibinfo{pages}{102001} (\bibinfo{year}{1999}).

\bibitem[{\citenamefont{Regimbau and Hughes}(2009)}]{prd.79.062002.09}
\bibinfo{author}{\bibfnamefont{T.}~\bibnamefont{Regimbau}} \bibnamefont{and}
  \bibinfo{author}{\bibfnamefont{S.~A.} \bibnamefont{Hughes}},
  \bibinfo{journal}{Phys. Rev. D} \textbf{\bibinfo{volume}{79}},
  \bibinfo{pages}{062002} (\bibinfo{year}{2009}).

\bibitem[{\citenamefont{Regimbau}(2011)}]{raap.11.369.11}
\bibinfo{author}{\bibfnamefont{T.}~\bibnamefont{Regimbau}},
  \bibinfo{journal}{Res. Astron. Astrophys.} \textbf{\bibinfo{volume}{11}},
  \bibinfo{pages}{369} (\bibinfo{year}{2011}).

\bibitem[{\citenamefont{Gair et~al.}(2011)\citenamefont{Gair, Mandel, Miller,
  and Volonteri}}]{grg.43.485.11}
\bibinfo{author}{\bibfnamefont{J.~R.} \bibnamefont{Gair}},
  \bibinfo{author}{\bibfnamefont{I.}~\bibnamefont{Mandel}},
  \bibinfo{author}{\bibfnamefont{M.~C.} \bibnamefont{Miller}},
  \bibnamefont{and}
  \bibinfo{author}{\bibfnamefont{M.}~\bibnamefont{Volonteri}},
  \bibinfo{journal}{Gen.Relativ. Gravit.} \textbf{\bibinfo{volume}{43}},
  \bibinfo{pages}{485} (\bibinfo{year}{2011}).

\bibitem[{\citenamefont{Belczynski et~al.}(2002)\citenamefont{Belczynski,
  Kalogera, and Bulik}}]{apj.572.407.02}
\bibinfo{author}{\bibfnamefont{K.}~\bibnamefont{Belczynski}},
  \bibinfo{author}{\bibfnamefont{V.}~\bibnamefont{Kalogera}}, \bibnamefont{and}
  \bibinfo{author}{\bibfnamefont{T.}~\bibnamefont{Bulik}},
  \bibinfo{journal}{Astrophys. J.} \textbf{\bibinfo{volume}{572}},
  \bibinfo{pages}{407} (\bibinfo{year}{2002}).

\bibitem[{\citenamefont{Belczynski et~al.}(2008)\citenamefont{Belczynski,
  Kalogera, Rasio, Taam, Zezas, Bulik, Maccarone, and
  Ivanova}}]{apjs.174.223.08}
\bibinfo{author}{\bibfnamefont{K.}~\bibnamefont{Belczynski}},
  \bibinfo{author}{\bibfnamefont{V.}~\bibnamefont{Kalogera}},
  \bibinfo{author}{\bibfnamefont{F.~A.} \bibnamefont{Rasio}},
  \bibinfo{author}{\bibfnamefont{R.~E.} \bibnamefont{Taam}},
  \bibinfo{author}{\bibfnamefont{A.}~\bibnamefont{Zezas}},
  \bibinfo{author}{\bibfnamefont{T.}~\bibnamefont{Bulik}},
  \bibinfo{author}{\bibfnamefont{T.~J.} \bibnamefont{Maccarone}},
  \bibnamefont{and} \bibinfo{author}{\bibfnamefont{N.}~\bibnamefont{Ivanova}},
  \bibinfo{journal}{Astrophys. J. Suppl.} \textbf{\bibinfo{volume}{174}},
  \bibinfo{pages}{223} (\bibinfo{year}{2008}).

\bibitem[{\citenamefont{Belczynski et~al.}(2010)\citenamefont{Belczynski,
  Dominik, Bulik, O'Shaughnessy, Fryer, and Holz}}]{apjl.715.L138.10}
\bibinfo{author}{\bibfnamefont{K.}~\bibnamefont{Belczynski}},
  \bibinfo{author}{\bibfnamefont{M.}~\bibnamefont{Dominik}},
  \bibinfo{author}{\bibfnamefont{T.}~\bibnamefont{Bulik}},
  \bibinfo{author}{\bibfnamefont{R.}~\bibnamefont{O'Shaughnessy}},
  \bibinfo{author}{\bibfnamefont{C.}~\bibnamefont{Fryer}}, \bibnamefont{and}
  \bibinfo{author}{\bibfnamefont{D.~E.} \bibnamefont{Holz}},
  \bibinfo{journal}{Astrophys. J. Lett.} \textbf{\bibinfo{volume}{715}},
  \bibinfo{pages}{L138} (\bibinfo{year}{2010}).

\bibitem[{\citenamefont{Dominik et~al.}(2012)\citenamefont{Dominik, Belczynski,
  Fryer, Holz, Berti, Bulik, Mandel, and O'Shaughnessy}}]{apj.759.52.12}
\bibinfo{author}{\bibfnamefont{M.}~\bibnamefont{Dominik}},
  \bibinfo{author}{\bibfnamefont{K.}~\bibnamefont{Belczynski}},
  \bibinfo{author}{\bibfnamefont{C.}~\bibnamefont{Fryer}},
  \bibinfo{author}{\bibfnamefont{D.~E.} \bibnamefont{Holz}},
  \bibinfo{author}{\bibfnamefont{E.}~\bibnamefont{Berti}},
  \bibinfo{author}{\bibfnamefont{T.}~\bibnamefont{Bulik}},
  \bibinfo{author}{\bibfnamefont{I.}~\bibnamefont{Mandel}}, \bibnamefont{and}
  \bibinfo{author}{\bibfnamefont{R.}~\bibnamefont{O'Shaughnessy}},
  \bibinfo{journal}{Astrophys. J.} \textbf{\bibinfo{volume}{759}},
  \bibinfo{pages}{52} (\bibinfo{year}{2012}).

\bibitem[{\citenamefont{Regimbau et~al.}(2014)\citenamefont{Regimbau, Meacher,
  and Coughlin}}]{prd.89.084046.14}
\bibinfo{author}{\bibfnamefont{T.}~\bibnamefont{Regimbau}},
  \bibinfo{author}{\bibfnamefont{D.}~\bibnamefont{Meacher}}, \bibnamefont{and}
  \bibinfo{author}{\bibfnamefont{M.}~\bibnamefont{Coughlin}},
  \bibinfo{journal}{Phys. Rev. D} \textbf{\bibinfo{volume}{89}},
  \bibinfo{pages}{084046} (\bibinfo{year}{2014}).

\bibitem[{\citenamefont{Cannon et~al.}(2010)\citenamefont{Cannon, Chapman,
  Hanna, Keppel, Searle, and Weinstein}}]{prd.82.044025.10}
\bibinfo{author}{\bibfnamefont{K.}~\bibnamefont{Cannon}},
  \bibinfo{author}{\bibfnamefont{A.}~\bibnamefont{Chapman}},
  \bibinfo{author}{\bibfnamefont{C.}~\bibnamefont{Hanna}},
  \bibinfo{author}{\bibfnamefont{D.}~\bibnamefont{Keppel}},
  \bibinfo{author}{\bibfnamefont{A.~C.} \bibnamefont{Searle}},
  \bibnamefont{and}
  \bibinfo{author}{\bibfnamefont{A.}~\bibnamefont{Weinstein}},
  \bibinfo{journal}{Phys. Rev. D} \textbf{\bibinfo{volume}{82}},
  \bibinfo{pages}{044025} (\bibinfo{year}{2010}).

\bibitem[{\citenamefont{Cannon et~al.}(2011)\citenamefont{Cannon, Hanna,
  Keppel, and Searle}}]{prd.83.084053.11}
\bibinfo{author}{\bibfnamefont{K.}~\bibnamefont{Cannon}},
  \bibinfo{author}{\bibfnamefont{C.}~\bibnamefont{Hanna}},
  \bibinfo{author}{\bibfnamefont{D.}~\bibnamefont{Keppel}}, \bibnamefont{and}
  \bibinfo{author}{\bibfnamefont{A.~C.} \bibnamefont{Searle}},
  \bibinfo{journal}{Phys. Rev. D} \textbf{\bibinfo{volume}{83}},
  \bibinfo{pages}{084053} (\bibinfo{year}{2011}).

\bibitem[{\citenamefont{Cannon et~al.}(2012)}]{apj.784.136.12}
\bibinfo{author}{\bibfnamefont{K.}~\bibnamefont{Cannon}} \bibnamefont{et~al.},
  \bibinfo{journal}{Astrophys. J.} \textbf{\bibinfo{volume}{748}},
  \bibinfo{pages}{136} (\bibinfo{year}{2012}).

\bibitem[{\citenamefont{Cannon et~al.}(2013)\citenamefont{Cannon, Hanna, and
  Keppel}}]{prd.88.024025.13}
\bibinfo{author}{\bibfnamefont{K.}~\bibnamefont{Cannon}},
  \bibinfo{author}{\bibfnamefont{C.}~\bibnamefont{Hanna}}, \bibnamefont{and}
  \bibinfo{author}{\bibfnamefont{D.}~\bibnamefont{Keppel}},
  \bibinfo{journal}{Phys. Rev. D} \textbf{\bibinfo{volume}{88}},
  \bibinfo{pages}{024025} (\bibinfo{year}{2013}).

\bibitem[{\citenamefont{Meacher et~al.}(2015)\citenamefont{Meacher, Coughlin,
  Morris, Regimbau, Christensen, Kandhasamy, Mandic, Romano, and
  Thrane}}]{prd.92.063002.15}
\bibinfo{author}{\bibfnamefont{D.}~\bibnamefont{Meacher}},
  \bibinfo{author}{\bibfnamefont{M.}~\bibnamefont{Coughlin}},
  \bibinfo{author}{\bibfnamefont{S.}~\bibnamefont{Morris}},
  \bibinfo{author}{\bibfnamefont{T.}~\bibnamefont{Regimbau}},
  \bibinfo{author}{\bibfnamefont{N.}~\bibnamefont{Christensen}},
  \bibinfo{author}{\bibfnamefont{S.}~\bibnamefont{Kandhasamy}},
  \bibinfo{author}{\bibfnamefont{V.}~\bibnamefont{Mandic}},
  \bibinfo{author}{\bibfnamefont{J.~D.} \bibnamefont{Romano}},
  \bibnamefont{and} \bibinfo{author}{\bibfnamefont{E.}~\bibnamefont{Thrane}},
  \bibinfo{journal}{Phys. Rev. D} \textbf{\bibinfo{volume}{92}},
  \bibinfo{pages}{063002} (\bibinfo{year}{2015}).

\bibitem[{\citenamefont{Freise et~al.}(2009)\citenamefont{Freise, Chelkowski,
  Hild, Del~Pozzo, Perreca, and Vecchio}}]{cqg.26.085012.09}
\bibinfo{author}{\bibfnamefont{A.}~\bibnamefont{Freise}},
  \bibinfo{author}{\bibfnamefont{S.}~\bibnamefont{Chelkowski}},
  \bibinfo{author}{\bibfnamefont{S.}~\bibnamefont{Hild}},
  \bibinfo{author}{\bibfnamefont{W.}~\bibnamefont{Del~Pozzo}},
  \bibinfo{author}{\bibfnamefont{A.}~\bibnamefont{Perreca}}, \bibnamefont{and}
  \bibinfo{author}{\bibfnamefont{A.}~\bibnamefont{Vecchio}},
  \bibinfo{journal}{Class. Quantum Grav.} \textbf{\bibinfo{volume}{26}},
  \bibinfo{pages}{085012} (\bibinfo{year}{2009}).

\bibitem[{\citenamefont{Punturo and Somiya}(2013)}]{ijmpd.22.1330010.13}
\bibinfo{author}{\bibfnamefont{M.}~\bibnamefont{Punturo}} \bibnamefont{and}
  \bibinfo{author}{\bibfnamefont{K.}~\bibnamefont{Somiya}},
  \bibinfo{journal}{Int. J. Mod. Phys. D} \textbf{\bibinfo{volume}{22}},
  \bibinfo{pages}{1330010} (\bibinfo{year}{2013}).

\bibitem[{\citenamefont{Thrane et~al.}(2013)\citenamefont{Thrane, Christensen,
  and Schofield}}]{prd.87.123009.13}
\bibinfo{author}{\bibfnamefont{E.}~\bibnamefont{Thrane}},
  \bibinfo{author}{\bibfnamefont{N.}~\bibnamefont{Christensen}},
  \bibnamefont{and} \bibinfo{author}{\bibfnamefont{R.~M.~S.}
  \bibnamefont{Schofield}}, \bibinfo{journal}{Phys. Rev. D}
  \textbf{\bibinfo{volume}{87}}, \bibinfo{pages}{123009}
  (\bibinfo{year}{2013}).

\bibitem[{\citenamefont{Thrane et~al.}(2014)\citenamefont{Thrane, Christensen,
  Schofield, and Effler}}]{prd.90.023013.14}
\bibinfo{author}{\bibfnamefont{E.}~\bibnamefont{Thrane}},
  \bibinfo{author}{\bibfnamefont{N.}~\bibnamefont{Christensen}},
  \bibinfo{author}{\bibfnamefont{R.~M.~S.} \bibnamefont{Schofield}},
  \bibnamefont{and} \bibinfo{author}{\bibfnamefont{A.}~\bibnamefont{Effler}},
  \bibinfo{journal}{Phys. Rev. D} \textbf{\bibinfo{volume}{90}},
  \bibinfo{pages}{023013} (\bibinfo{year}{2014}).

\bibitem[{\citenamefont{Hopkins and Beacom}(2006)}]{apj.651.142.06}
\bibinfo{author}{\bibfnamefont{A.~M.} \bibnamefont{Hopkins}} \bibnamefont{and}
  \bibinfo{author}{\bibfnamefont{J.~F.} \bibnamefont{Beacom}},
  \bibinfo{journal}{Astrophys. J.} \textbf{\bibinfo{volume}{651}},
  \bibinfo{pages}{142} (\bibinfo{year}{2006}).

\bibitem[{\citenamefont{Abadie et~al.}(2010{\natexlab{b}})}]{cqg.27.173001.10}
\bibinfo{author}{\bibfnamefont{J.}~\bibnamefont{Abadie}} \bibnamefont{et~al.}
  (\bibinfo{collaboration}{LIGO Scientific Collaboration and Virgo
  Collaboration}), \bibinfo{journal}{Class. Quantum Grav.}
  \textbf{\bibinfo{volume}{27}}, \bibinfo{pages}{173001}
  (\bibinfo{year}{2010}{\natexlab{b}}).

\bibitem[{\citenamefont{Kowalska et~al.}(2015)\citenamefont{Kowalska, Regimbau,
  Bulik, Dominik, and Belczynski}}]{aap.574.A58.15}
\bibinfo{author}{\bibfnamefont{I.}~\bibnamefont{Kowalska}},
  \bibinfo{author}{\bibfnamefont{T.}~\bibnamefont{Regimbau}},
  \bibinfo{author}{\bibfnamefont{T.}~\bibnamefont{Bulik}},
  \bibinfo{author}{\bibfnamefont{M.}~\bibnamefont{Dominik}}, \bibnamefont{and}
  \bibinfo{author}{\bibfnamefont{K.}~\bibnamefont{Belczynski}},
  \bibinfo{journal}{Astron. \& Astrophys.} \textbf{\bibinfo{volume}{574}},
  \bibinfo{pages}{A58} (\bibinfo{year}{2015}).

\bibitem[{\citenamefont{Buonanno et~al.}(2009)\citenamefont{Buonanno, Iyer,
  Ochsner, Pan, and Sathyaprakash}}]{prd.80.084043.09}
\bibinfo{author}{\bibfnamefont{A.}~\bibnamefont{Buonanno}},
  \bibinfo{author}{\bibfnamefont{B.~R.} \bibnamefont{Iyer}},
  \bibinfo{author}{\bibfnamefont{E.}~\bibnamefont{Ochsner}},
  \bibinfo{author}{\bibfnamefont{Y.}~\bibnamefont{Pan}}, \bibnamefont{and}
  \bibinfo{author}{\bibfnamefont{B.~S.} \bibnamefont{Sathyaprakash}},
  \bibinfo{journal}{Phys. Rev. D} \textbf{\bibinfo{volume}{80}},
  \bibinfo{pages}{084043} (\bibinfo{year}{2009}).

\bibitem[{\citenamefont{Blanchet}(2014)}]{lrr.17.2.14}
\bibinfo{author}{\bibfnamefont{L.}~\bibnamefont{Blanchet}},
  \bibinfo{journal}{Living Rev. Relativity} \textbf{\bibinfo{volume}{17}},
  \bibinfo{pages}{2} (\bibinfo{year}{2014}).

\bibitem[{\citenamefont{Brown et~al.}(2013)\citenamefont{Brown, Kumar, and
  Nitz}}]{prd.87.082004.13}
\bibinfo{author}{\bibfnamefont{D.~A.} \bibnamefont{Brown}},
  \bibinfo{author}{\bibfnamefont{P.}~\bibnamefont{Kumar}}, \bibnamefont{and}
  \bibinfo{author}{\bibfnamefont{A.~H.} \bibnamefont{Nitz}},
  \bibinfo{journal}{Phys. Rev. D} \textbf{\bibinfo{volume}{87}},
  \bibinfo{pages}{082004} (\bibinfo{year}{2013}).

\bibitem[{\citenamefont{Damour et~al.}(2000)\citenamefont{Damour, Iyer, and
  Sathyaprakash}}]{prd.62.084036.00}
\bibinfo{author}{\bibfnamefont{T.}~\bibnamefont{Damour}},
  \bibinfo{author}{\bibfnamefont{B.~R.} \bibnamefont{Iyer}}, \bibnamefont{and}
  \bibinfo{author}{\bibfnamefont{B.~S.} \bibnamefont{Sathyaprakash}},
  \bibinfo{journal}{Phys. Rev. D} \textbf{\bibinfo{volume}{62}},
  \bibinfo{pages}{084036} (\bibinfo{year}{2000}).

\bibitem[{\citenamefont{Welch}(1967)}]{itae.15.70.67}
\bibinfo{author}{\bibfnamefont{P.~D.} \bibnamefont{Welch}},
  \bibinfo{journal}{IEEE Trans. Audio Electroacoust}
  \textbf{\bibinfo{volume}{15}}, \bibinfo{pages}{70} (\bibinfo{year}{1967}).

\bibitem[{\citenamefont{Allen et~al.}(2012)\citenamefont{Allen, Anderson,
  Brady, Brown, and Creighton}}]{prd.85.122006.12}
\bibinfo{author}{\bibfnamefont{B.}~\bibnamefont{Allen}},
  \bibinfo{author}{\bibfnamefont{W.~G.} \bibnamefont{Anderson}},
  \bibinfo{author}{\bibfnamefont{P.~R.} \bibnamefont{Brady}},
  \bibinfo{author}{\bibfnamefont{D.~A.} \bibnamefont{Brown}}, \bibnamefont{and}
  \bibinfo{author}{\bibfnamefont{J.~D.~E.} \bibnamefont{Creighton}},
  \bibinfo{journal}{Phys. Rev. D} \textbf{\bibinfo{volume}{85}},
  \bibinfo{pages}{122006} (\bibinfo{year}{2012}).

\bibitem[{\citenamefont{Aasi et~al.}(2013{\natexlab{a}})}]{ObsScenario}
\bibinfo{author}{\bibfnamefont{J.}~\bibnamefont{Aasi}} \bibnamefont{et~al.}
  (\bibinfo{collaboration}{LIGO Scientific Collaboration and Virgo
  Collaboration}), \bibinfo{journal}{arXiv} \textbf{\bibinfo{volume}{1304}},
  \bibinfo{pages}{0670} (\bibinfo{year}{2013}{\natexlab{a}}).

\bibitem[{\citenamefont{Brady and Fairhurst}(2008)}]{cqg.25.105002.08}
\bibinfo{author}{\bibfnamefont{P.~R.} \bibnamefont{Brady}} \bibnamefont{and}
  \bibinfo{author}{\bibfnamefont{S.}~\bibnamefont{Fairhurst}},
  \bibinfo{journal}{Class. Quantum Grav.} \textbf{\bibinfo{volume}{25}},
  \bibinfo{pages}{105002} (\bibinfo{year}{2008}).

\bibitem[{\citenamefont{Van Den~Broeck}(2014)}]{jpcs.484.012008.14}
\bibinfo{author}{\bibfnamefont{C.}~\bibnamefont{Van Den~Broeck}},
  \bibinfo{journal}{J. Phys.: Conf. Ser.} \textbf{\bibinfo{volume}{484}},
  \bibinfo{pages}{012008} (\bibinfo{year}{2014}).

\bibitem[{\citenamefont{Aasi et~al.}(2014{\natexlab{a}})}]{apj.785.119.14}
\bibinfo{author}{\bibfnamefont{J.}~\bibnamefont{Aasi}} \bibnamefont{et~al.}
  (\bibinfo{collaboration}{LIGO Scientific Collaboration and Virgo
  Collaboration}), \bibinfo{journal}{Astrophys. J.}
  \textbf{\bibinfo{volume}{785}}, \bibinfo{pages}{119}
  (\bibinfo{year}{2014}{\natexlab{a}}).

\bibitem[{\citenamefont{Aasi et~al.}(2013{\natexlab{b}})}]{prd.88.102002.13}
\bibinfo{author}{\bibfnamefont{J.}~\bibnamefont{Aasi}} \bibnamefont{et~al.}
  (\bibinfo{collaboration}{LIGO Scientific Collaboration and Virgo
  Collaboration}), \bibinfo{journal}{Phys. Rev. D}
  \textbf{\bibinfo{volume}{88}}, \bibinfo{pages}{102002}
  (\bibinfo{year}{2013}{\natexlab{b}}).

\bibitem[{\citenamefont{Aasi et~al.}(2014{\natexlab{b}})}]{prd.90.062010.14}
\bibinfo{author}{\bibfnamefont{J.}~\bibnamefont{Aasi}} \bibnamefont{et~al.}
  (\bibinfo{collaboration}{LIGO Scientific Collaboration and Virgo
  Collaboration}), \bibinfo{journal}{Phys. Rev. D}
  \textbf{\bibinfo{volume}{90}}, \bibinfo{pages}{062010}
  (\bibinfo{year}{2014}{\natexlab{b}}).

\bibitem[{\citenamefont{Bin{\'e}truy et~al.}(2012)\citenamefont{Bin{\'e}truy,
  Boh{\'e}, Caprini, and Dufaux}}]{jcap.6.27.12}
\bibinfo{author}{\bibfnamefont{P.}~\bibnamefont{Bin{\'e}truy}},
  \bibinfo{author}{\bibfnamefont{A.}~\bibnamefont{Boh{\'e}}},
  \bibinfo{author}{\bibfnamefont{C.}~\bibnamefont{Caprini}}, \bibnamefont{and}
  \bibinfo{author}{\bibfnamefont{J.-F.} \bibnamefont{Dufaux}},
  \bibinfo{journal}{Journal of Cosmology and Astroparticle Physics}
  \textbf{\bibinfo{volume}{6}}, \bibinfo{pages}{27} (\bibinfo{year}{2012}).

\bibitem[{\citenamefont{Caprini et~al.}(2008)\citenamefont{Caprini, Durrer, and
  Servant}}]{prd.77.124015.08}
\bibinfo{author}{\bibfnamefont{C.}~\bibnamefont{Caprini}},
  \bibinfo{author}{\bibfnamefont{R.}~\bibnamefont{Durrer}}, \bibnamefont{and}
  \bibinfo{author}{\bibfnamefont{G.}~\bibnamefont{Servant}},
  \bibinfo{journal}{Phys. Rev. D} \textbf{\bibinfo{volume}{77}},
  \bibinfo{pages}{124015} (\bibinfo{year}{2008}).

\bibitem[{\citenamefont{Caprini
  et~al.}(2009{\natexlab{a}})\citenamefont{Caprini, Durrer, Konstandin, and
  Servant}}]{prd.79.083519.09}
\bibinfo{author}{\bibfnamefont{C.}~\bibnamefont{Caprini}},
  \bibinfo{author}{\bibfnamefont{R.}~\bibnamefont{Durrer}},
  \bibinfo{author}{\bibfnamefont{T.}~\bibnamefont{Konstandin}},
  \bibnamefont{and} \bibinfo{author}{\bibfnamefont{G.}~\bibnamefont{Servant}},
  \bibinfo{journal}{Phys. Rev. D} \textbf{\bibinfo{volume}{79}},
  \bibinfo{pages}{083519} (\bibinfo{year}{2009}{\natexlab{a}}).

\bibitem[{\citenamefont{Caprini
  et~al.}(2009{\natexlab{b}})\citenamefont{Caprini, Durrer, and
  Servant}}]{jcap.12.024.09}
\bibinfo{author}{\bibfnamefont{C.}~\bibnamefont{Caprini}},
  \bibinfo{author}{\bibfnamefont{R.}~\bibnamefont{Durrer}}, \bibnamefont{and}
  \bibinfo{author}{\bibfnamefont{G.}~\bibnamefont{Servant}},
  \bibinfo{journal}{Journal of Cosmology and Astroparticle Physics}
  \textbf{\bibinfo{volume}{12}}, \bibinfo{pages}{024}
  (\bibinfo{year}{2009}{\natexlab{b}}).

\bibitem[{\citenamefont{Damour and Vilenkin}(2005)}]{prd.71.063510.05}
\bibinfo{author}{\bibfnamefont{T.}~\bibnamefont{Damour}} \bibnamefont{and}
  \bibinfo{author}{\bibfnamefont{A.}~\bibnamefont{Vilenkin}},
  \bibinfo{journal}{Phys. Rev. D} \textbf{\bibinfo{volume}{71}},
  \bibinfo{pages}{063510} (\bibinfo{year}{2005}).

\bibitem[{\citenamefont{Siemens et~al.}(2007)\citenamefont{Siemens, Mandic, and
  Creighton}}]{prl.98.111101.07}
\bibinfo{author}{\bibfnamefont{X.}~\bibnamefont{Siemens}},
  \bibinfo{author}{\bibfnamefont{V.}~\bibnamefont{Mandic}}, \bibnamefont{and}
  \bibinfo{author}{\bibfnamefont{J.}~\bibnamefont{Creighton}},
  \bibinfo{journal}{Phys. Rev. Lett.} \textbf{\bibinfo{volume}{98}},
  \bibinfo{pages}{111101} (\bibinfo{year}{2007}).

\bibitem[{\citenamefont{{\"O}lmez et~al.}(2010)\citenamefont{{\"O}lmez, Mandic,
  and Siemens}}]{prd.81.104028.10}
\bibinfo{author}{\bibfnamefont{S.}~\bibnamefont{{\"O}lmez}},
  \bibinfo{author}{\bibfnamefont{V.}~\bibnamefont{Mandic}}, \bibnamefont{and}
  \bibinfo{author}{\bibfnamefont{X.}~\bibnamefont{Siemens}},
  \bibinfo{journal}{Phys. Rev. D} \textbf{\bibinfo{volume}{81}},
  \bibinfo{pages}{104028} (\bibinfo{year}{2010}).

\bibitem[{\citenamefont{Regimbau
  et~al.}(2012{\natexlab{b}})\citenamefont{Regimbau, Giampanis, Siemens, and
  Mandic}}]{prd.85.066001.12}
\bibinfo{author}{\bibfnamefont{T.}~\bibnamefont{Regimbau}},
  \bibinfo{author}{\bibfnamefont{S.}~\bibnamefont{Giampanis}},
  \bibinfo{author}{\bibfnamefont{X.}~\bibnamefont{Siemens}}, \bibnamefont{and}
  \bibinfo{author}{\bibfnamefont{V.}~\bibnamefont{Mandic}},
  \bibinfo{journal}{Phys. Rev. D} \textbf{\bibinfo{volume}{85}},
  \bibinfo{pages}{066001} (\bibinfo{year}{2012}{\natexlab{b}}).

\bibitem[{\citenamefont{Aasi et~al.}(2014{\natexlab{c}})}]{prl.112.131101.13}
\bibinfo{author}{\bibfnamefont{J.}~\bibnamefont{Aasi}} \bibnamefont{et~al.}
  (\bibinfo{collaboration}{LIGO Scientific Collaboration and Virgo
  Collaboration}), \bibinfo{journal}{Phys. Rev. Lett.}
  \textbf{\bibinfo{volume}{112}}, \bibinfo{pages}{131101}
  (\bibinfo{year}{2014}{\natexlab{c}}).

\bibitem[{\citenamefont{Gasperini and Veneziano}(1993)}]{app.1.317.93}
\bibinfo{author}{\bibfnamefont{M.}~\bibnamefont{Gasperini}} \bibnamefont{and}
  \bibinfo{author}{\bibfnamefont{G.}~\bibnamefont{Veneziano}},
  \bibinfo{journal}{Astropart. Phys.} \textbf{\bibinfo{volume}{1}},
  \bibinfo{pages}{317} (\bibinfo{year}{1993}).

\bibitem[{\citenamefont{Buonanno et~al.}(1997)\citenamefont{Buonanno, Maggiore,
  and Ungarelli}}]{prd.55.3330.97}
\bibinfo{author}{\bibfnamefont{A.}~\bibnamefont{Buonanno}},
  \bibinfo{author}{\bibfnamefont{M.}~\bibnamefont{Maggiore}}, \bibnamefont{and}
  \bibinfo{author}{\bibfnamefont{C.}~\bibnamefont{Ungarelli}},
  \bibinfo{journal}{Phys. Rev. D} \textbf{\bibinfo{volume}{55}},
  \bibinfo{pages}{3330} (\bibinfo{year}{1997}).

\bibitem[{\citenamefont{Dufaux et~al.}(2010)\citenamefont{Dufaux, Figueroa, and
  Garc{\'{\i}}a-Bellido}}]{prd.82.083518.10}
\bibinfo{author}{\bibfnamefont{J.~F.} \bibnamefont{Dufaux}},
  \bibinfo{author}{\bibfnamefont{D.~G.} \bibnamefont{Figueroa}},
  \bibnamefont{and}
  \bibinfo{author}{\bibfnamefont{J.}~\bibnamefont{Garc{\'{\i}}a-Bellido}},
  \bibinfo{journal}{Phys. Rev. D} \textbf{\bibinfo{volume}{82}},
  \bibinfo{pages}{083518} (\bibinfo{year}{2010}).

\bibitem[{\citenamefont{Wu et~al.}(2012)\citenamefont{Wu, Mandic, and
  Regimbau}}]{prd.85.104024.12}
\bibinfo{author}{\bibfnamefont{C.}~\bibnamefont{Wu}},
  \bibinfo{author}{\bibfnamefont{V.}~\bibnamefont{Mandic}}, \bibnamefont{and}
  \bibinfo{author}{\bibfnamefont{T.}~\bibnamefont{Regimbau}},
  \bibinfo{journal}{Phys. Rev. D} \textbf{\bibinfo{volume}{85}},
  \bibinfo{pages}{104024} (\bibinfo{year}{2012}).

\bibitem[{\citenamefont{Buonanno et~al.}(05)\citenamefont{Buonanno, Chen, Pan,
  Tagoshi, and Vallisneri}}]{prd.72.084027.05}
\bibinfo{author}{\bibfnamefont{A.}~\bibnamefont{Buonanno}},
  \bibinfo{author}{\bibfnamefont{Y.}~\bibnamefont{Chen}},
  \bibinfo{author}{\bibfnamefont{Y.}~\bibnamefont{Pan}},
  \bibinfo{author}{\bibfnamefont{H.}~\bibnamefont{Tagoshi}}, \bibnamefont{and}
  \bibinfo{author}{\bibfnamefont{M.}~\bibnamefont{Vallisneri}},
  \bibinfo{journal}{Phys. Rev. D} \textbf{\bibinfo{volume}{72}},
  \bibinfo{pages}{084027} (\bibinfo{year}{05}).

\bibitem[{\citenamefont{Buonanno
  et~al.}(2006{\natexlab{a}})\citenamefont{Buonanno, Chen, and
  Vallisneri}}]{prd.67.104025.06}
\bibinfo{author}{\bibfnamefont{A.}~\bibnamefont{Buonanno}},
  \bibinfo{author}{\bibfnamefont{Y.}~\bibnamefont{Chen}}, \bibnamefont{and}
  \bibinfo{author}{\bibfnamefont{M.}~\bibnamefont{Vallisneri}},
  \bibinfo{journal}{Phys. Rev. D} \textbf{\bibinfo{volume}{67}},
  \bibinfo{pages}{104025} (\bibinfo{year}{2006}{\natexlab{a}}).

\bibitem[{\citenamefont{Buonanno
  et~al.}(2006{\natexlab{b}})\citenamefont{Buonanno, Chen, Pan, and
  Vallisneri}}]{prd.74.029902.06}
\bibinfo{author}{\bibfnamefont{A.}~\bibnamefont{Buonanno}},
  \bibinfo{author}{\bibfnamefont{Y.}~\bibnamefont{Chen}},
  \bibinfo{author}{\bibfnamefont{Y.}~\bibnamefont{Pan}}, \bibnamefont{and}
  \bibinfo{author}{\bibfnamefont{M.}~\bibnamefont{Vallisneri}},
  \bibinfo{journal}{Phys. Rev. D} \textbf{\bibinfo{volume}{74}},
  \bibinfo{pages}{029902} (\bibinfo{year}{2006}{\natexlab{b}}).

\bibitem[{\citenamefont{\'{E}anna \'{E}.~Flanagan and
  Hinderer}(2008)}]{prd.77.021502.08}
\bibinfo{author}{\bibnamefont{\'{E}anna \'{E}.~Flanagan}} \bibnamefont{and}
  \bibinfo{author}{\bibfnamefont{T.}~\bibnamefont{Hinderer}},
  \bibinfo{journal}{Phys. Rev. D} \textbf{\bibinfo{volume}{77}},
  \bibinfo{pages}{021502} (\bibinfo{year}{2008}).

\bibitem[{\citenamefont{Read et~al.}(2009)\citenamefont{Read, Markakis,
  Shibata, Uryu, Creighton, and Friedman}}]{prd.79.124033.09}
\bibinfo{author}{\bibfnamefont{J.~S.} \bibnamefont{Read}},
  \bibinfo{author}{\bibfnamefont{C.}~\bibnamefont{Markakis}},
  \bibinfo{author}{\bibfnamefont{M.}~\bibnamefont{Shibata}},
  \bibinfo{author}{\bibfnamefont{K.}~\bibnamefont{Uryu}},
  \bibinfo{author}{\bibfnamefont{J.~D.~E.} \bibnamefont{Creighton}},
  \bibnamefont{and} \bibinfo{author}{\bibfnamefont{J.~L.}
  \bibnamefont{Friedman}}, \bibinfo{journal}{Phys. Rev. D}
  \textbf{\bibinfo{volume}{79}}, \bibinfo{pages}{124033}
  (\bibinfo{year}{2009}).

\bibitem[{\citenamefont{Hinderer et~al.}(2010)\citenamefont{Hinderer, Lackey,
  Lang, and Read}}]{prd.81.123016.10}
\bibinfo{author}{\bibfnamefont{T.}~\bibnamefont{Hinderer}},
  \bibinfo{author}{\bibfnamefont{B.~D.} \bibnamefont{Lackey}},
  \bibinfo{author}{\bibfnamefont{R.~N.} \bibnamefont{Lang}}, \bibnamefont{and}
  \bibinfo{author}{\bibfnamefont{J.~S.} \bibnamefont{Read}},
  \bibinfo{journal}{Phys. Rev. D} \textbf{\bibinfo{volume}{81}},
  \bibinfo{pages}{123016} (\bibinfo{year}{2010}).

\bibitem[{\citenamefont{Pannarale et~al.}(2011)\citenamefont{Pannarale,
  Rezzolla, Ohme, and Read}}]{prd.84.104017.11}
\bibinfo{author}{\bibfnamefont{F.}~\bibnamefont{Pannarale}},
  \bibinfo{author}{\bibfnamefont{L.}~\bibnamefont{Rezzolla}},
  \bibinfo{author}{\bibfnamefont{F.}~\bibnamefont{Ohme}}, \bibnamefont{and}
  \bibinfo{author}{\bibfnamefont{J.~S.} \bibnamefont{Read}},
  \bibinfo{journal}{Phys. Rev. D} \textbf{\bibinfo{volume}{84}},
  \bibinfo{pages}{104017} (\bibinfo{year}{2011}).

\bibitem[{\citenamefont{Hannam et~al.}(2014)\citenamefont{Hannam, Schmidt,
  Boh\'{e}, Haegel, Husa, Ohme, Pratten, and P\"{u}rrer}}]{prl.113.151101.14}
\bibinfo{author}{\bibfnamefont{M.}~\bibnamefont{Hannam}},
  \bibinfo{author}{\bibfnamefont{P.}~\bibnamefont{Schmidt}},
  \bibinfo{author}{\bibfnamefont{A.}~\bibnamefont{Boh\'{e}}},
  \bibinfo{author}{\bibfnamefont{L.}~\bibnamefont{Haegel}},
  \bibinfo{author}{\bibfnamefont{S.}~\bibnamefont{Husa}},
  \bibinfo{author}{\bibfnamefont{F.}~\bibnamefont{Ohme}},
  \bibinfo{author}{\bibfnamefont{G.}~\bibnamefont{Pratten}}, \bibnamefont{and}
  \bibinfo{author}{\bibfnamefont{M.}~\bibnamefont{P\"{u}rrer}},
  \bibinfo{journal}{Phys. Rev. Lett.} \textbf{\bibinfo{volume}{113}},
  \bibinfo{pages}{151101} (\bibinfo{year}{2014}).

\bibitem[{\citenamefont{Schmidt et~al.}(2015)\citenamefont{Schmidt, Ohme, and
  Hannam}}]{prd.91.024043.15}
\bibinfo{author}{\bibfnamefont{P.}~\bibnamefont{Schmidt}},
  \bibinfo{author}{\bibfnamefont{F.}~\bibnamefont{Ohme}}, \bibnamefont{and}
  \bibinfo{author}{\bibfnamefont{M.}~\bibnamefont{Hannam}},
  \bibinfo{journal}{Phys. Rev. D} \textbf{\bibinfo{volume}{91}},
  \bibinfo{pages}{024043} (\bibinfo{year}{2015}).

\bibitem[{\citenamefont{Barr et~al.}(2012)}]{StrawmanRed}
\bibinfo{author}{\bibfnamefont{B.}~\bibnamefont{Barr}} \bibnamefont{et~al.},
  \bibinfo{journal}{LIGO DCC}  (\bibinfo{year}{2012}),
  \urlprefix\url{https://dcc.ligo.org/LIGO-T1200046/public}.

\bibitem[{\citenamefont{Nissanke et~al.}(2010)\citenamefont{Nissanke, Holz,
  Hughes, Dalal, and Sievers}}]{apj.725.496.10}
\bibinfo{author}{\bibfnamefont{S.}~\bibnamefont{Nissanke}},
  \bibinfo{author}{\bibfnamefont{D.~E.} \bibnamefont{Holz}},
  \bibinfo{author}{\bibfnamefont{S.~A.} \bibnamefont{Hughes}},
  \bibinfo{author}{\bibfnamefont{N.}~\bibnamefont{Dalal}}, \bibnamefont{and}
  \bibinfo{author}{\bibfnamefont{J.~L.} \bibnamefont{Sievers}},
  \bibinfo{journal}{Astrophys. J.} \textbf{\bibinfo{volume}{725}},
  \bibinfo{pages}{496} (\bibinfo{year}{2010}).

\bibitem[{\citenamefont{Pozzo}(2012)}]{prd.86.043011.12}
\bibinfo{author}{\bibfnamefont{W.~D.} \bibnamefont{Pozzo}},
  \bibinfo{journal}{Phys. Rev. D} \textbf{\bibinfo{volume}{86}},
  \bibinfo{pages}{043011} (\bibinfo{year}{2012}).

\bibitem[{\citenamefont{Taylor and Gair}(2012)}]{prd.86.023502.12}
\bibinfo{author}{\bibfnamefont{S.~R.} \bibnamefont{Taylor}} \bibnamefont{and}
  \bibinfo{author}{\bibfnamefont{J.~R.} \bibnamefont{Gair}},
  \bibinfo{journal}{Phys. Rev. D} \textbf{\bibinfo{volume}{86}},
  \bibinfo{pages}{023502} (\bibinfo{year}{2012}).

\bibitem[{\citenamefont{Messenger and Read}(2012)}]{prl.108.091101.12}
\bibinfo{author}{\bibfnamefont{C.}~\bibnamefont{Messenger}} \bibnamefont{and}
  \bibinfo{author}{\bibfnamefont{J.}~\bibnamefont{Read}},
  \bibinfo{journal}{Phys. Rev. Lett.} \textbf{\bibinfo{volume}{108}},
  \bibinfo{pages}{091101} (\bibinfo{year}{2012}).

\end{thebibliography}

\end{document}